\newcommand{\pv}{\ensuremath{p_V}}
\newcommand{\pV}{\ensuremath{p_V}}

\renewcommand{\deg}{\ensuremath{^\circ}}

\newcommand{\gcc}{g cm$^{-3}$}

\newcommand{\hide}[1]{} 

\documentclass[11pt,preprint]{aastex} 

\usepackage[squaren,textstyle]{SIunits}
\usepackage{url}

\usepackage{amssymb}

\begin{document}

\title{Introducing the Eulalia and new Polana asteroid families:
  re-assessing primitive asteroid families in the inner Main Belt}

\author{Kevin J. Walsh} \affil{Southwest Research Institute, 1050
  Walnut St. Suite 400, Boulder, CO, 80302, USA}
\email{kwalsh@boulder.swri.edu}

\author{Marco Delb\'{o}} \affil{UNS-CNRS-Observatoire de la C\^ote
  d'Azur, BP 4229, 06304 Nice cedex 04, France}

\author{William F. Bottke}\affil{Southwest Research Institute, 1050
  Walnut St. Suite 400, Boulder, CO, 80302, USA}

\author{David Vokrouhlick\'{y}}\affil{Institute of Astronomy, Charles
  University, V Hole\v{s}ovi\v{c}k\'{a}ch 2, 18000 Prague 8, Czech
  Republic}

\author{Dante S. Lauretta}\affil{Lunar \& Planetary Laboratory, University of Arizona, Tucson, AZ 85721, USA}

\begin{abstract}
  The so-called Nysa-Polana complex of asteroids is a diverse and
  widespread group studied by Cellino et al. (2001, 2002) as a
  dynamically linked asteroid family. It carries the name of two
  asteroids because it appears to be two overlapping families of
  different asteroid taxonomies: (44) Nysa is an E-type asteroid with
  the lowest number in the midst of a predominantly S-type cluster and
  (142) Polana is a B-type asteroid near the low-albedo B- and C-type
  cluster. The latter has been shown to be a very important source of
  primitive Near Earth Asteroids.

  Using the data from the Wide-field Infrared Survey Explorer (WISE)
  mission we have re-analyzed the region around the Nysa-Polana
  complex in the inner Main Belt, focusing on the low-albedo
  population. (142) Polana does not appear to be a member of the
  family of low-albedo asteroids in the Nysa-Polana complex. Rather,
  the largest is asteroid (495) Eulalia. This asteroid has never before
  been linked to this complex for an important dynamical reason: it
  currently has a proper eccentricity slightly below the range of most
  of the family members. However, its orbit is very close to the 3:1
  mean motion resonance with Jupiter and is in a weak secular
  resonance. We show that its osculating eccentricity varies widely
  ($e=0.06-0.19$) on short timescales ($\sim$1~Myr) and the averaged
  value diffuses (between $e=0.11-0.15$) over long timescales
  ($\sim$100~Myr). The diffusive orbit, low-albedo, taxonomic
  similarity and semimajor axis strongly suggests that despite its
  current proper eccentricity, (495) Eulalia could have recently been
  at an orbit very central to the family. Hierarchical Clustering
  Method tests confirm that at an eccentricity of $e=$0.15, (495)
  Eulalia could be the parent of the family.  The ``Eulalia family''
  was formed between 900--1500~Myr ago, and likely resulted from the
  breakup of a 100--160~km parent body.

  There is also compelling evidence for an older and more widespread
  primitive family in the same region of the asteroid belt parented by
  asteroid (142) Polana. This family, the ``new Polana family'', is
  more extended in orbital elements, and is older than 2000~Myr.

\end{abstract}


\section{Introduction}

\subsection{NEO origins}

Near-Earth objects (NEOs) are temporary visitors in the region around
the terrestrial planets, and significant work has gone into
understanding their dynamical behavior and lifetimes. Gladman et
al. (2000) and others found that NEOs have short lifetimes, only about
10~Myr on average.  Gravitational perturbations from planets,
collisions with the terrestrial planets, or loss into the Sun or
  ejection out of the Solar System limit their average lifetimes.
Contrasted with the age of the Solar System, the much shorter NEO
lifetimes suggest that today's NEOs are simply the current incarnation
of a constantly re-filled steady-state population. Cratering records
(Grieve and Shoemaker 1994; St\"{o}ffler and Ryder 2001) find that the
impact flux on the terrestrial planets and the Moon has been
relatively constant on Gyr timescales, supporting the idea of a
steady-state population.

The problem of how to refill NEO space from the seemingly static Main
Asteroid belt was answered with the discovery that asteroids are quite
mobile (Bottke et al. 2000, 2006). Thermal forces work via the
Yarkovsky-effect to change asteroid's semimajor axes over time
(Farinella and Vokrouhlick\'{y} 1999; Bottke et al. 2000, 2001; see
also a review by Bottke et al. 2006). This discovery provided the key
piece of physics needed to understand how Main Belt asteroids become
NEOs. It had been known that some of the more powerful resonances
located near, or in, the Main Asteroid belt could rapidly excite
asteroid's eccentricities, sending them onto planet-crossing orbits
leading to their delivery to NEO orbits. The Yarkovsky effect showed
how to efficiently get asteroids into these resonances.

The two most important resonances for this delivery process are
readily visible in a plot of the asteroid belt. First, the 3:1 mean
motion resonance (MMR) with Jupiter is located at heliocentric
  distance $\sim$2.5~AU, and is responsible for the large gap in the
asteroid distribution at this semi-major axis. The other is the
$\nu_{6}$ secular resonance, which occurs when the precession
frequency of an asteroid's longitude of perihelion is equal to the
mean precession frequency of Saturn. This resonance is inclination
dependent, and is very efficient at delivering low-inclination bodies
to NEO orbits. It is the effective inner edge of the asteroid belt at
$\sim$2.15~AU, and estimated to deliver $\sim$37\% of all NEOs with
$H<18$ (Bottke et al. 2002).

The inner Main Belt (IMB; $2.15 < a < 2.5$~AU) -- bound by these two
resonances -- is a predominant source of NEOs. Dynamical models
predict that $\sim$61\% of the $H < 18$ NEO population comes from
there. The majority of the detected IMB asteroids,
  $\sim$~4/5 those with $H < 15.5$, are on low-inclination orbits
  ($i < 8^\circ$). Although a compositional
gradient is known to exist in the main belt, with low-albedo,
primitive asteroids being predominant in the central (2.5~AU$ < a <
$2.8~AU) and the outer asteroid belt ($a > $2.8~AU) (Gradie and
Tedesco 1978; Moth\'{e}-Diniz et al. 2003; Masiero et al. 2011),
 the IMB contains numerous
primitive asteroids (Campins et al. 2010; Gayon-Markt et al. 2012;
Masiero et al. 2011). For
instance, in a sample of WISE-studied asteroids limited to absolute
magnitude $H<15$, about 1/6 of these bodies in the IMB with measured
albedos and sizes have geometric visible albedos \pv$<0.1$, where we
use \pv$<0.1$ as a simple way to separate low-albedo primitive bodies
from more processed or igneous bodies typically with higher albedos.

Recent studies devoted to finding the origin of the primitive NEOs
1999 RQ$_{36}$, 1999 JU$_{3}$, and 1996 FG$_{3}$ (baseline targets of the
sample return space missions, OSIRIS-REx, Hayabusa-II, and Marco
Polo-R) have found that each of these three bodies are almost
certainly ($>$90\%) delivered from the IMB following the well-studied
dynamical pathway from the Main Belt to NEO-orbits (Campins et
al. 2010, 2012; Walsh et al. 2012; Bottke et al. 2000,
2002). Moreover, the current low inclination ($i<$ 8\deg) orbits of
these bodies is indicative of origins on similarly low inclination
orbits in the Main Belt. Furthermore, Jenniskens et al. (2010), and
Gayon-Markt et al. (2012) have idenified
the IMB at low inclination ($i<$ 8\deg) as the likely source ($>$90\%)
of the NEO 2008 TC$_3$, the asteroid whose impact produced the
Almahata Sitta meteorites.

The work presented here develops the fundamental hypotheses to be
tested by planned asteroid sample return missions. Dynamical evolution
studies are critical to obtaining the maximum scientific benefit from
these missions. Such studies support primary mission objectives to
characterize the geologic and dynamic history of the target asteroids
and provide critical context for their returned samples. In addition,
cosmochemical analyses of e.g., cosmogenic isotope ratios,
radionuclide abundances, and nuclear track densities will, in return,
provide important constraints on the dynamical evolution of the parent
asteroid. This synergy will result in improved understanding of the
dynamical pathways that transform IMB objects into NEOs. Furthermore,
the OSIRIS-REx mission specifically will provide the first
ground-truth assessment of the Yarkovsky effect as it relates to the
chemical nature and dynamical state of an individual asteroid. These
studies will provide important input to models of the evolution of
asteroid families by Yarkovsky drift and delivery to orbital
resonances in the IMB.

\subsection{Asteroid families and the Nysa-Polana complex}

\begin{figure}[h!]
\includegraphics[angle=-90,width=\linewidth]{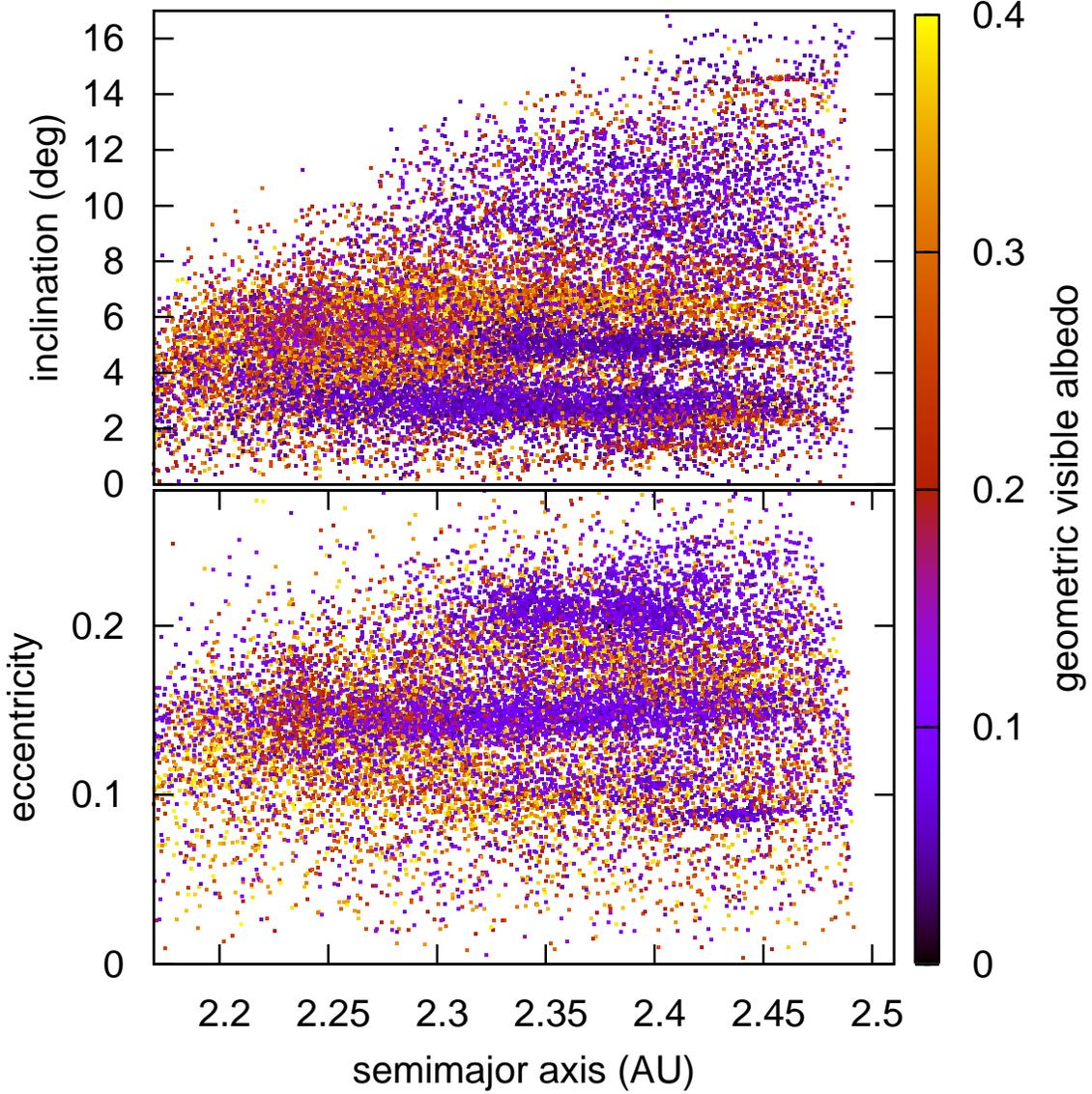}
\caption{The entire inner Main Belt, including all asteroids that were
  observed by WISE and whose $D$ and \pv\ were published in the
  Preliminary release of Albedos and Diameters (Masiero et al. 2011)
  plotted using computed synthetic proper elements (Kne\v{z}evi\'{c}
  and Milani 2003). The plots show
  the orbital inclination and the eccentricity as a function of their
  semimajor axis (AU). The color of each point represents their
  WISE-determined albedo with values shown in the colorbar on the
  right (Masiero et al. 2011). The low-albedo component of the
  Nysa-Polana complex is visible as the large low-albedo complex at
  $e\sim0.15$ and $i\sim3^\circ$ extending across almost the entire
  IMB.  \label{fig:IMBalbedo}}
\end{figure}

The landscape of the inner Main Belt is dominated by a handful of
asteroid families and a diffuse population of ``background'' objects
(see Fig. \ref{fig:IMBalbedo}).  Among primitive, low-albedo, bodies
the largest known family is the low-albedo component of the
Nysa-Polana complex, followed by the Erigone family and the Sulamitis
family (Nesvorn\'{y} 2010; Gayon-Markt
et al. 2012; Campins et al. 2012). Following an asteroid break-up
event, fragments are launched onto orbits that are distinct from, but
similar to, the parent body. Specifically, their ``proper orbital
elements'', roughly the long-term average of their osculating orbital
elements, remain linked over time (Zappal\`{a} et al. 1990, 1994;
Kne\v{z}evi\'{c} and Milani 2003).  Smaller fragments are typically
launched with higher velocity, creating a size-dependent spread in
orbital elements. Over time the thermal Yarkovsky effect induces a
size-dependent drift in semi-major axis, such that smaller bodies
drift faster, and thus further, over time than the larger
bodies. Therefore families of asteroids are identifiable in two ways:
clustering in proper orbital element space, and correlated shapes in
size vs. semimajor axis due to size-dependent Yarkovsky drift (see
Vokrouhlick\'y et al. 2006a and Bottke et al. 2006).

In the context of NEO-delivery, the size, age and location of an
asteroid family are important properties. First, the larger the
family, the more potential asteroids that can be delivered. Second,
the age of a family determines how far it has spread by the
Yarkovsky-effect, where older families can spread further. Finally,
the location of the family determines how far its fragments must drift
via the Yarkovsky effect to reach a resonance. As we find with the
Nysa-Polana complex, a location very near a resonance makes locating
the center and determining the age of the family much more difficult.

Not all asteroids are associated with a specific asteroid family.
There are a similar amount of ``background'' objects in this region as
there are members of an identified family (Gayon-Markt et
al. 2012). This work will, however, focus on asteroid families for a
few reasons. First, asteroid families imply a physical correlation
between fragments. Thus, we may learn more about an interesting
asteroid by studying a much larger and brighter parent body or a suite
of associated family members. Second, families' orbital elements
evolve in a relatively simple way - objects drift in semimajor axis at
a speed relative to the inverse of their size due to the Yarkovsky
effect. This means that a family member that has drifted into a
resonance can be followed by equal or smaller-sized members of the
same family at the same time - delivery of objects can be correlated
temporally.

The specific asteroid family targeted in this work is part of the
larger Nysa-Polana complex (NPC) - which appears to be two overlapping
asteroid families of different asteroid taxonomies. Their overlapping
nature has made detailed study difficult, but now the large database
of albedos measured by the WISE mission has made a study of each of
the two componenets of the family possible (Masiero et al. 2011;
Mainzer et al. 2011). 

Section \ref{s:familyidentification} describes the analysis of the
families in the low-albedo component of the Nysa-Polana complex -
seeking to find the family center, parent and age. Section
\ref{sec:Properties} describes the known physical properties of the
family and Section \ref{discussion} analyzes a second primitive family
in the same region.

\section{Identification of the Eulalia family}
\label{s:familyidentification}

As suggested in works by Zappal\`{a} et al. (1995), Cellino et
al. (2001, 2002), and Nesvorn\'{y} (2010) one component of the
Nysa-Polana complex consists of low-albedo objects belonging to C- or
B-type taxonomies (see also Campins et al. 2010).  This work will
re-analyze the low-albedo population of the inner Main Belt (at low
inclinations and moderate eccentricities) in order to find the center,
the parent and the age of any low-albedo families in this region.  The
first target is the low-albedo component of the Nysa-Polana complex
seeking to constrain the center and find the parent of the family,
followed by an age estimate.

\subsection{Data sources}
\label{s:datasources}

We have used the calculated albedos from Masiero et
al. (2011) to study asteroids observed by the NASA WISE
mission in the IMB with low albedo. The WISE mission measured fluxes
in 4 wavelengths 3.4, 4.6, 12 and 22 $\mu$m, and combined with a NEATM
thermal model calculated visible albedos, (\pv), and diameters ($D$)
for over 100,000 Main Belt asteroids. Not all asteroids were observed
and not all observed asteroids were detected the same number of
times. Therefore errors on albedo and diameter can vary from object to
object, and also the sample of objects is not complete at any
diameter. However, this work is not aiming to build a definitive list
of family members, rather the aims are to better understand the family
structure in the inner Main Belt. This database gives  accurate
measurements for a representative sample of bodies.

Within this dataset there are 6702 asteroids in the inner Main Belt
($2.1$~AU$ < a < 2.5$~AU) that are measured to have albedo, \pv$ <
$~0.1, by WISE and that also have synthetic proper orbital elements in
the database of 318,112 objects as of May 2012 from the {\tt AstDyS}
database {\tt http://hamilton.dm.unipi.it/astdys} (see also
Kne\v{z}evi\'{c} and Milani 2003).  The Nysa-Polana complex (NPC) of
asteroids is centered around $e\sim 0.15$ and $i\sim 3^{\circ}$. Hence
out of the entire IMB, we take a subset with orbital eccentricities
$0.1 < e < 0.2$ with $i<10^\circ$.  This area is centered on the NPC,
but contains a large portion of the entire IMB. The largest $H$
  value in this dataset is $H=18.8$.

\begin{figure}[h!]
\includegraphics[angle=0,width=\linewidth]{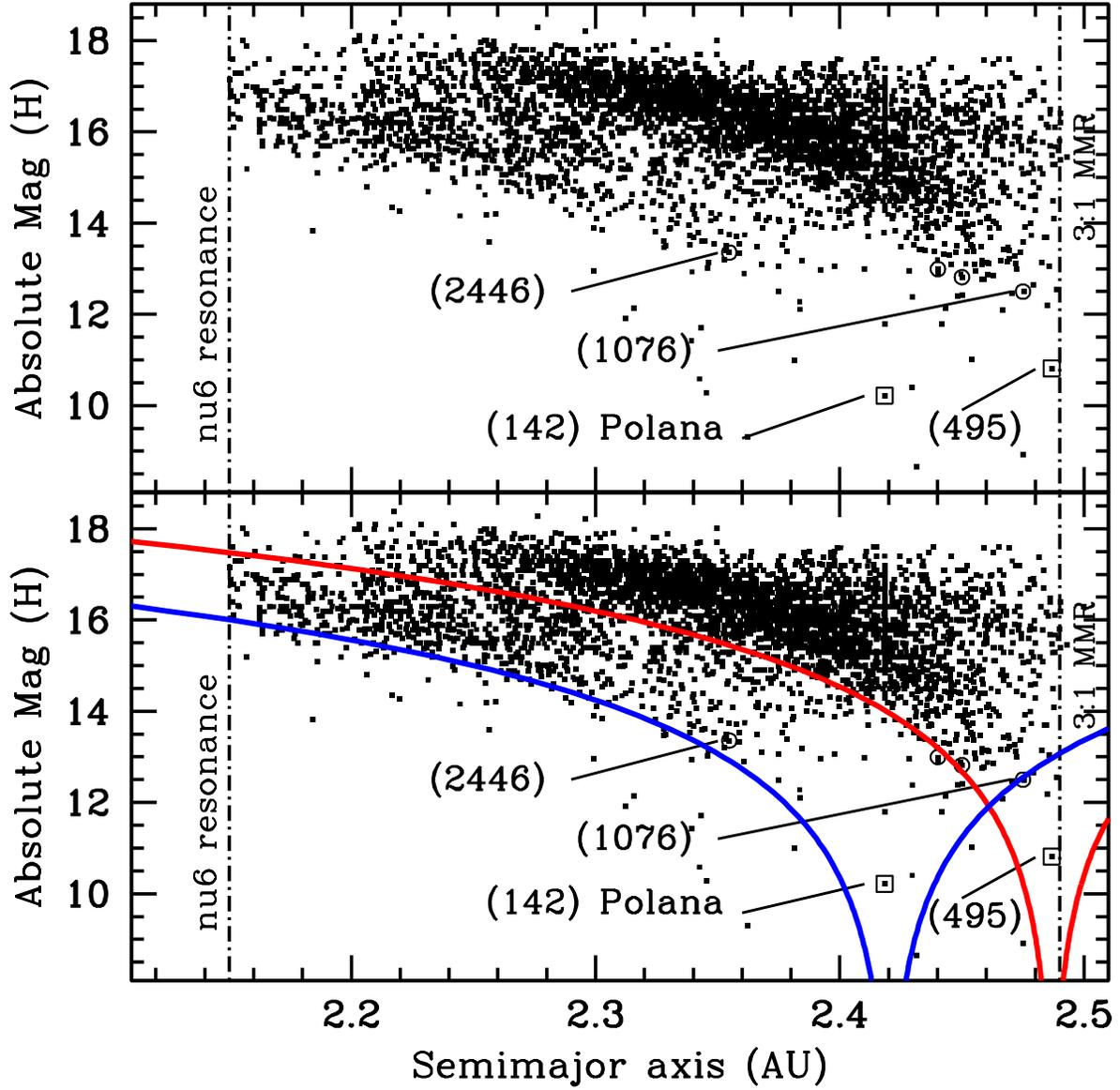}
\caption{The absolute magnitude, $H$, for all WISE-observed asteroids
  with albedo \pv\ below 0.1, eccentricities between 0.1 and 0.2 and
  inclinations below 10$^\circ$, plotted as a function of their
  semi-major axes (AU). The points with circles around their dots
  indicate the objects with published visible and near-IR spectra. The
  bottom panel is the same as the top, except lines are drawn to
  outline the Yarkovsky ``V-shape'' in the data. \label{fig:Vshape} }
\end{figure}

This subset of asteroids has a distinct ``V-shape'', or rather half of
a V-shape, in the distribution of the absolute magnitude $H$ as a
function of $a$ (Fig.~\ref{fig:Vshape}; see also Gayon-Markt et
al. 2012). This V-shape is typical of asteroid families due to the
size dependence of the Yarkovsky effect semimajor axis drift of the
family members (see Vokrouhlick\'{y} et al. 2006a or Bottke et
al. 2006). This is expected since the selection of asteroids
  surrounding the Nysa-Polana complex should be dominated by
  low-albedo family members that are expected to show a correlation in
  semimajor axis and absolute magnitude. However, two half V-shapes
  are visible, indicating the possible presence of two families with
  asteroids of low albedo in the region. This section is focused on
the higher density V-shape structure, with the other analyzed in Section
\ref{discussion}.

\subsection{The center of the low-albedo component of the Nysa-Polana
  complex}

Identification and analysis of the low-albedo families in the inner
Main Asteroid belt pose a few unique problems. First, half of the
typical asteroid family ``V''-shape is visible in
Fig. \ref{fig:Vshape}, which suggests that the parent body is
very close to the 3:1 mean motion resonance with Jupiter. So close, in
fact, that we can assume that at least half of the family was
immediately lost to the powerful resonance shortly after the formation
of the family (though we do search for family members on the other
side of the resonance).  Second, the families are very extended,
nearly reaching from the 3:1 to the $\nu_6$, suggesting an old age and
increasing the possibilities of interlopers and confusion in
determining family membership. Lastly, there may be more than one
family.

Typically, asteroid families are centered on a known parent body as
determined by Hierarchical Cluster Method (HCM) fitting. With a known
center the next variable to match is the age of the family which is
determined by the width of the bounding envelope due to the
size-dependent Yarkovsky drift of the asteroids. However, in this
case, the shape of the asteroid distribution was not aligned with the
previously proposed parent asteroid (142) Polana.

Given the uncertainty in both the center and boundaries of the family,
we seek to constrain these one at a time. First, following
Vokrouhlick\'{y} et al. (2006a), the $C$ parameter can be used to normalize asteroid distance from a possible family center by
their size. The Yarkovsky drift rate is inversely proportional to an
asteroid's diameter, so removing this measureable allows a search for
clustering of drift distances from possible asteroid family
centers. The $C$ parameter is defined
\begin{equation}
C=\Delta a \times 10^{-H/5}
\end{equation}
where $H$ is the absolute magnitude of the asteroid and $\Delta a =
\frac{da}{dt}T$ is the distance an asteroid has drifted from its
initial semimajor axis ($a_{init}$) over a time $T$ at a rate
$da/dt$. This formulation of $C$ normalizes each asteroid by its size,
as the Yarkovsky drift rate is inversely proportional to its size
$da/dt \propto D^{-1}$ such that $da/dt = (da/dt)_0(D_0/D)$. For a
simplistic asteroid family formation event where all bodies start with
the same $a_{init}$ and similar spin axes, all family members will
have the same $C$, since this value normalizes their drift rate by
their size. This is simplified by ignoring both the spin-axis
evolution over time and different $a_{init}$ for each asteroid due to
the dispersion of fragments following the family-forming impact.

\begin{figure}[h!]
\includegraphics[width=\linewidth]{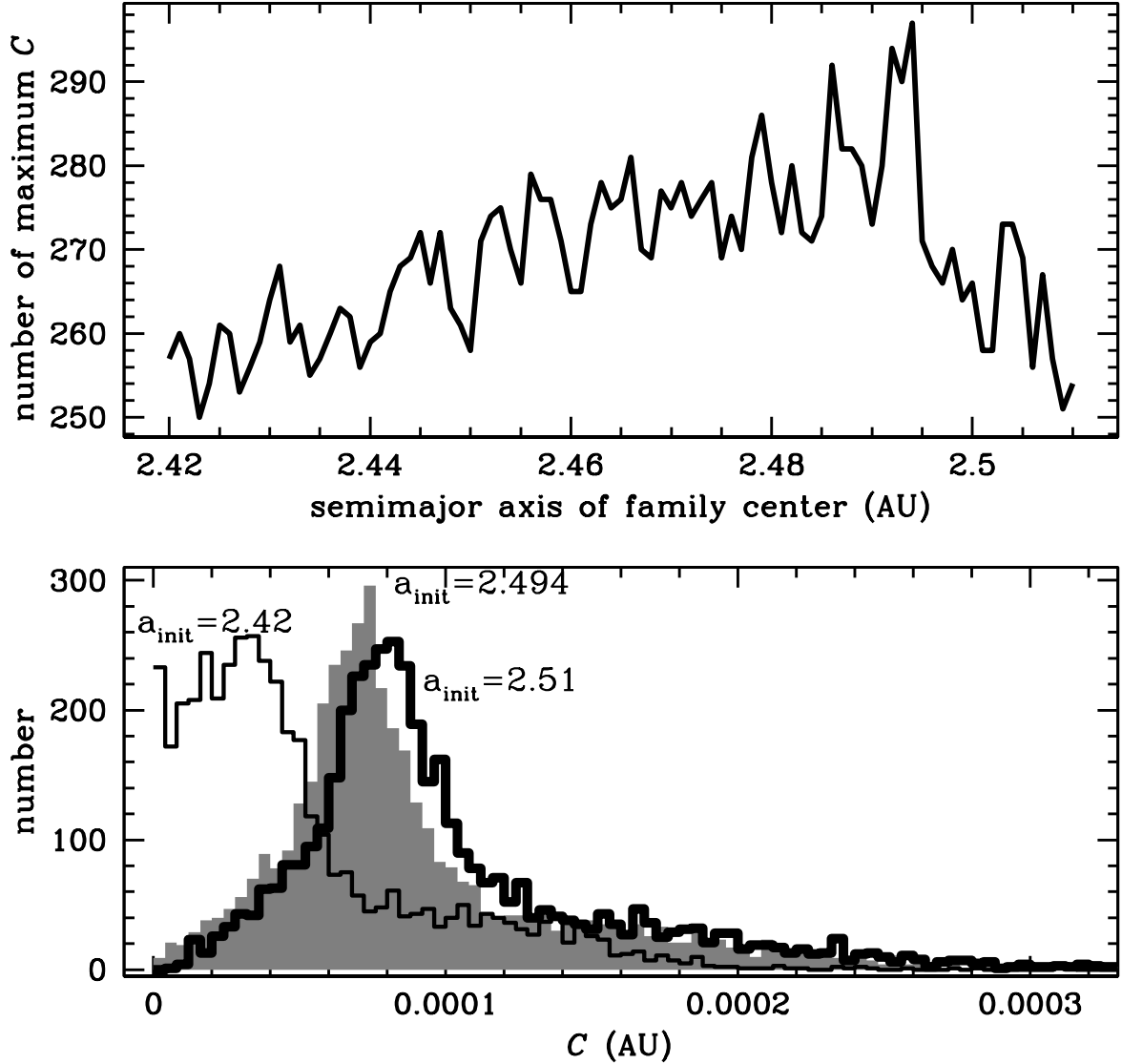}
\caption{ The top panel shows the number of bodies in the maximum bin
  for the $C$ value as a function of a center semimajor axis value
  ($a_{init}$).
  The bottom panel shows a histogram of $C$ values for all
  asteroids in the subset of WISE data ($2.15$~AU $< a <2.5$~AU ; $0.1
  < e < 0.2$; $i < 10^{\circ}$), for the best-fit family center
  location of $a=2.494$~AU. The bin size is $\Delta
  C=4\times10^{-6}$~AU, and the maximum is found at
  $C=7.4\times10^{-5}$~AU.  \label{fig:CenSearch}} 
\end{figure}

However, the $C$ parameter can help search for the center of the
family. The distribution of $C$ for a given $a_{init}$ would be flat
for a random distribution of asteroids. If a family is present that
had correlated $C$ values due to their common semimajor axis drift
times, then for an $a_{init}$ near the center of the family there will
be a spike in $C$ values. Thus, the $C$ distribution can be
re-calculated at different $a_{init}$, seeking the value that provide
the tightest clustering of $C$. The search algorithm simply counted
the number of asteroids in the maximum bin of $\Delta
C=4\times10^{-6}$~AU, and the range of $a_{init}$ was selected to
  cover the centers suggested by Fig. \ref{fig:Vshape}. As $a_{init}$
was varied from $a=$2.42--2.51~AU, a clear trend was observed, with a
peak at $a=2.494$~AU (Fig. \ref{fig:CenSearch}).

\subsection{The diffusive orbit of (495) Eulalia}

The previous section found the center of the family is best fit at
$a\sim2.494$~AU, which is very close to the 3:1 MMR with Jupiter
nominally located at $\sim$~2.5~AU. In the selected subset of
asteroids there is only one asteroid with $2.48$~AU $< a < 2.5$~AU
with an absolute magnitude less than 12 and only three with $H$ less
than 13.  We have also searched known asteroids with proper
elements in the same region, finding only a few with measured low
albedo from IRAS or AKARI (asteroids 877 and 2159, Tedesco et
al. 2002, Usui et al. 2011), but both have inclinations around
3.5$^\circ$, outside the boundaries of the estimated family. The only
asteroid with suitable inclination, semimajor axis and absolute
magnitude is (495)~Eulalia, with $H=10.8$ an albedo \pv=0.051 and a
diameter $D\sim40$~km (Masiero et al. 2011). The synthetic proper
orbital elements are $a=2.4868$~AU, $e=0.1184, \sin i =0.0438$
(Kne\v{z}evi\'{c} and Milani 2003). The synthetic proper eccentricity
of $e=0.118$ is well below the NPC family average, which is why it was
never previously identified as a candidate parent for any of the
Nysa-Polana complex.

However, the orbit of (495) Eulalia is extremely close the the 3:1
mean motion resonance (MMR) with Jupiter. The location of this
resonance is nominally at $\sim$~2.5~AU where a body has an
orbital period 1/3 that of Jupiter. The location of the resonance is
also a function of the small body's orbital eccentricity, where larger
eccentricities expands the location of the resonance in semimajor axis
(see Wisdom 1993, 1995). The boundary of the resonance itself is not
precise, rather the chaoticity of an orbit increases as it approaches
the resonance. Therefore the precise behavior of (495) Eulalia over
time cannot be estimated trivially (see also Guillens et al. 2002).

To estimate the chaotic affects on (495) Eulalia due to its location
near the 3:1~MMR with Jupiter, its long-term evolution was
modeled. Based on the published uncertainties for its current
osculating orbital elements, we integrated 1000 clone particles
(created from the 6D hyper-ellipsoid in the osculating elements space
compatible with today's orbit). The clones were integrated using the
{\tt swift\_mvs} code, which is part of the {\tt SWIFT} integration
package (Levison and Duncan 1994). All 8 planets were included in the
simulation, and each clone was treated as a test particle (feeling
only the gravity of the planets). The simulations spanned 500~Myr
(with some tests to 4 Gyr).

\begin{figure}[h!]
\includegraphics[angle=0,width=\linewidth]{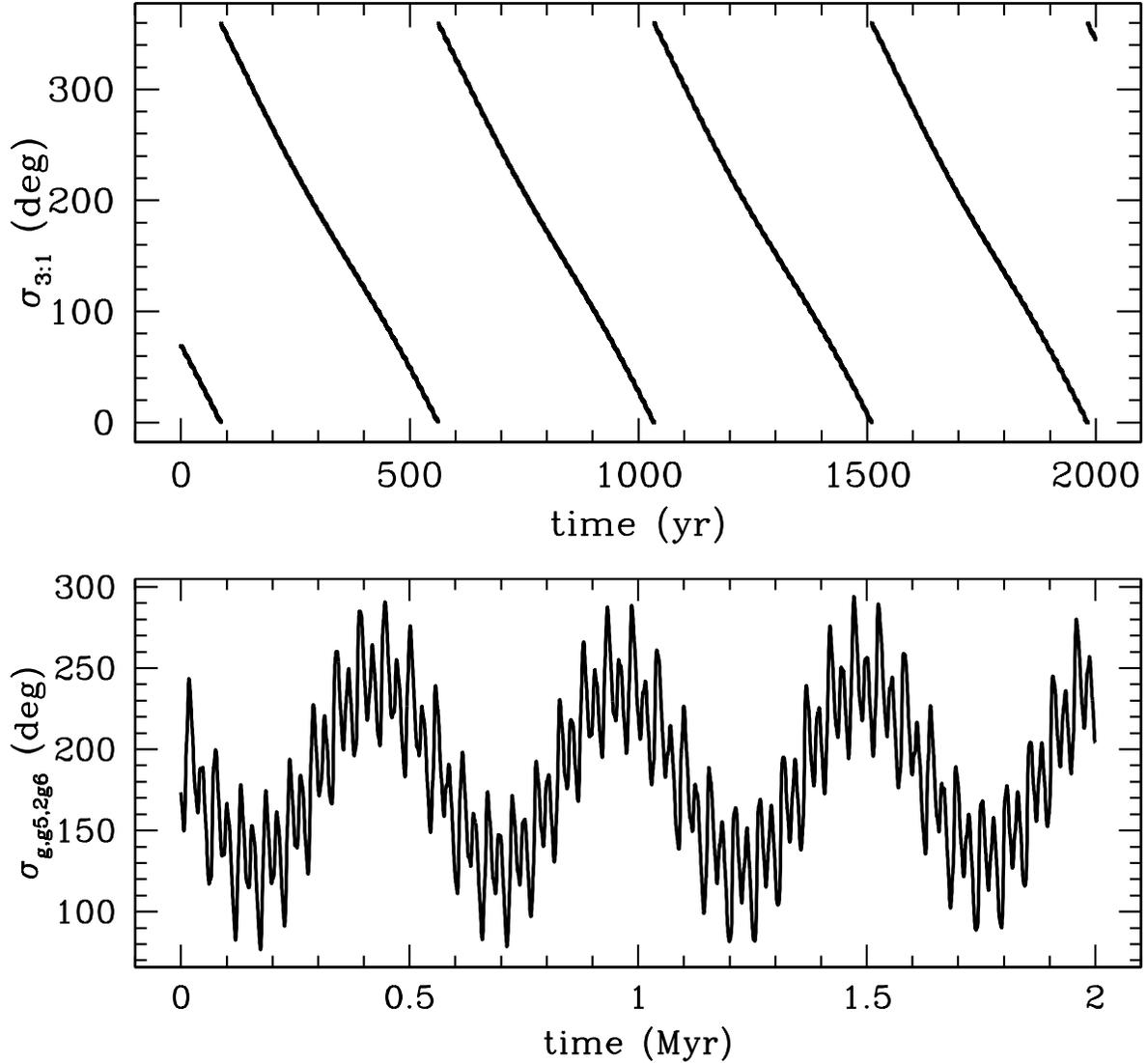}
\caption{Eulalia's orbital interactions with the 3:1 MMR with Jupiter
  and also a secular resonance $g+g_5-2g_6$ . The top panel shows the
  slow (relative to the orbital periods) circulation of the principal
  resonance angle $\sigma_{31} =
  3\lambda_{\mathrm{Jupiter}}-\lambda-2\varpi$. The
  bottom panel shows the much longer timescale interaction with the
  $g_5$ and $g_6$ frequencies with the argument
  $\sigma_{g,g5,2g6}=\varpi+\varpi_{\rm Jupiter}-2\varpi_{\rm Saturn}$
  librating about $180^\circ$ with a period of $\sim 600$~ky (where short-period oscillations were eliminated by digital filtering). }
\label{fig:resangle}
\end{figure}

The specific behavior for any single particle at very short timescales
(kyr) is dominated by the proximity to the 3:1 mean motion resonance
(MMR) with Jupiter via the slow circulation of the principal resonance
angle $\sigma = 3\lambda_{\mathrm{Jupiter}}-\lambda-2\varpi$
(where $\lambda$ is the orbital frequency of an orbit and $\varpi$
  is the precession frequency of an orbit; see
  Fig. \ref{fig:resangle}). On longer timescales the interaction with
$g+g_5-2g_6$ secular resonance results in the corresponding
resonance angle $\sigma=\varpi+\varpi_{\rm Jupiter}-2\varpi_{\rm
  Saturn}$ librating about $180^\circ$ with a period of $\sim 600$~ky
(where $g$, $g_5$ and $g_6$ are the proper frequencies associated
  with the precession of the asteroid, Jupiter and Saturn
  respectively).  This interaction, together with other secular
  terms of the asteroid's orbit, is responsible for osculating
eccentricity oscillations between 0.06--0.19.

\begin{figure}[h!]
\includegraphics[angle=0,width=\linewidth]{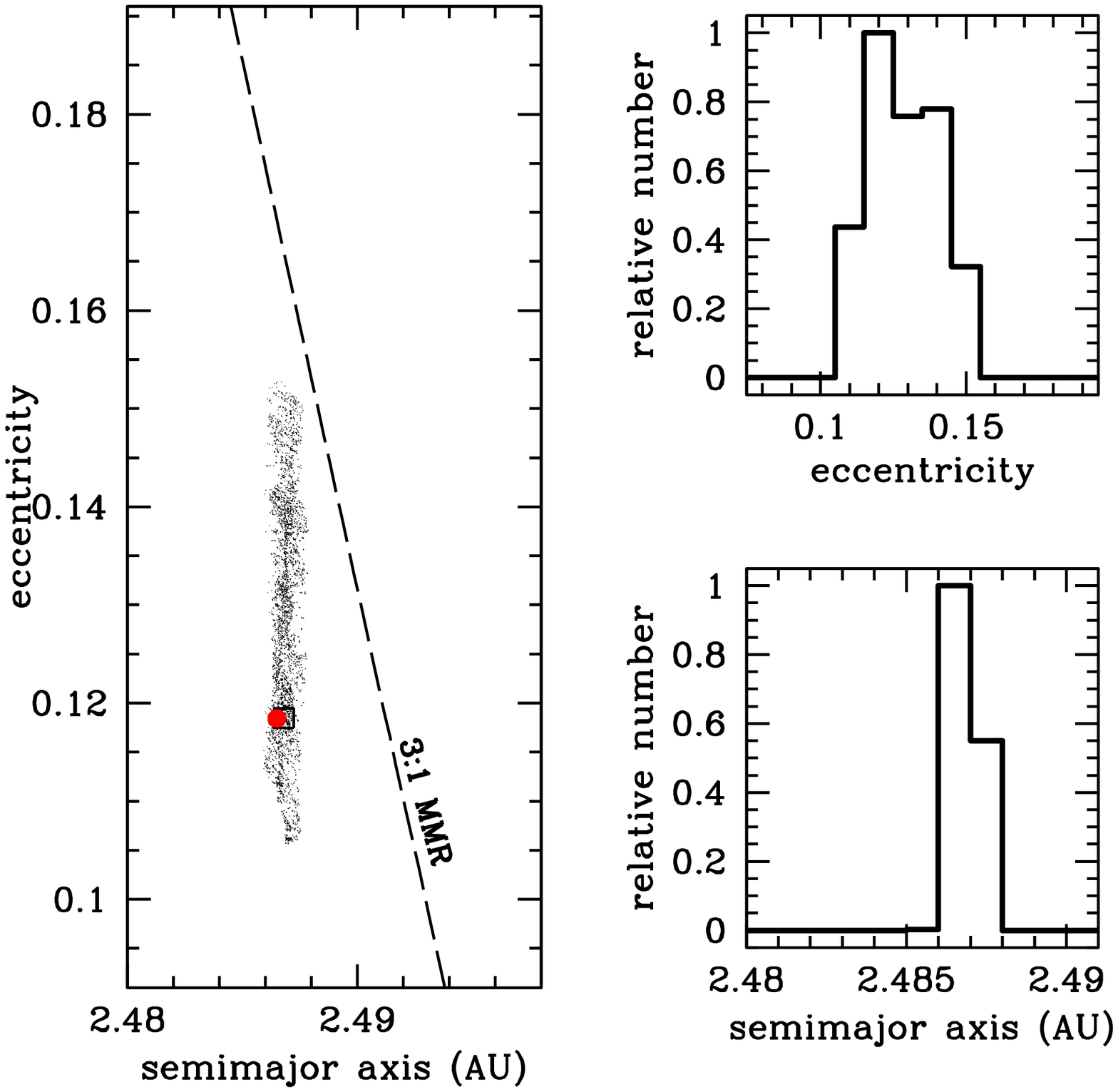}
\caption{The evolution for 500~Myr of a subset of 200~clones of
  (495) Eulalia.  The large red dot shows the starting location for a
  single clone and also the current synthetic proper elements of
  (495) Eulalia, the black dots show the evolution of the orbit averaged over
  10~Myr timescales.  The dashed line is the nominal location of the
  3:1~MMR separatrix plotted as eccentricity as a function of
  semimajor axis (Nesvorn\'{y} et al. 2002). The upper right
  panel shows the time spent by each of the 200 clones at different
  values of 10~Myr averaged eccentricity. The bottom right panel
  shows the same for semimajor axis. On longer timescales,
  during 1--2~Gyr, the $e=0.15$ bin is occupied $\sim$~13\% of the
  time.}
\label{fig:diffuse}
\end{figure}

To analyze long-term stability and orbital evolution the entire suite
of particles was considered with averaged orbital elements over 10~Myr
time periods as a proxy for proper elements. Each particle's averaged
elements was sampled every 10~Myr, creating a distribution mapping the
time spent at different orbits - specifically at different
eccentricities (Fig. \ref{fig:diffuse}). The semimajor axis of the 1000
clones is quite constant with $a\simeq 2.486$--$2.487$~AU, while the average
eccentricity diffuses between $e$=0.11--0.15, with most of its time
spent at $e$=0.12. From the limited tests run to 4~Gyr we find some
clones are lost to the 3:1~MMR; at 2~Gyr only 30\% of the clones are
lost and at 4~Gyr only about 60\% of the clones are lost. So despite
being chaotic, the gravitational-only orbit of (495) Eulalia is
surprisingly stable for long timescales.

The role of thermal effects on the dynamics of Eulalia may be
important given the putative old age of the family. At
$D\sim40$~km its maximum drift rate over 1~Gyr is around 0.005~AU,
which is larger than the diffusion found in semimajor axis in the long
timescale integrations above. Also, given its precarious location very
near the 3:1~MMR, a drift of 0.005~AU could change its long-term
evolution dramatically, including a slightly different orbit at
  the time of a possible family-forming impact. A more realistic
estimate of its potential drift rate depends heavily on its obliquity
and various thermal properties. Its rotation rate and obliquity have
been estimated at 29.2$\pm$1.0~hr and $\sim$88$^\circ$ (Binzel
1987). Increasing the accuracy of these measurements and estimating
possible Yarkovsky drift scenarios, along with the possible role of
the YORP effect and secular spin-orbit resonances, is beyond the scope
of this work.

The conclusion of this test is that when searching for family
membership using an HCM model, the proper eccentricity of (495)
Eulalia should be tested at values between $e$=0.11--0.15, as they are
all possible given the asteroid'
s proximity to the 3:1~MMR.

\subsection{Hierarchical Clustering Method Family Search around (495)
  Eulalia}

If (495) Eulalia is the parent of a family caused by a collision, the
family-forming collision could have happened when the asteroid had any
of its allowable orbital eccentricities. Therefore a Hierarchical
Clustering Method search for possible family members associated with
(495) Eulalia, necessarily needs to be done varying its proper
eccentricity\footnote{Note that we have also tested and verified
  family formation during a maximum of the secular eccentricity
  oscillation.}.

Family association was tested using the Hierarchical Clustering Method
(see Bendjoya and Zappal\`{a} 2002 and references therein). HCM
searches among asteroid's proper orbital elements, as it is assumed
that proper orbital elements are stable on very long timescales, with
the exception of the drift in semimajor axis due to the Yarkovsky
affect. The algorithm relies on the ``standard metric'' of Zappal\`{a} et
al. (1990, 1995) to calculate a velocity difference between two
asteroid orbits. It then connects bodies falling within a cut-off
velocity ($V_c$), where $V_c$ is typically $<100$~m/s. 

However, the region surrounding (495) Eulalia is particularly
low-density, presumably due to its location directly adjacent to the
3:1~MMR. Even small Yarkovsky drifts to higher semimajor axis would
drive an asteroid directly into the resonance, thus Eulalia presumably
lost many of its closest neighbors long ago. The HCM search here
relied on the the WISE-surveyed low-albedo asteroids, and has only
6702 objects in the database, as compared to over 100,000 IMB
asteroids with synthetic proper elements. So, where typically a $V_c >
100$~m/s is at risk of linking to a substantial fraction of the
asteroid belt, here even $V_c \sim 200$~m/s still only links asteroids
with similar orbits.  With (495) Eulalia as the parent, with its $0.11
< e <0.16$, HCM tests were run varying $V_c$ between $40 < V_c <
200$~m/s.

\begin{figure}[h!]
\includegraphics[angle=0,width=\linewidth]{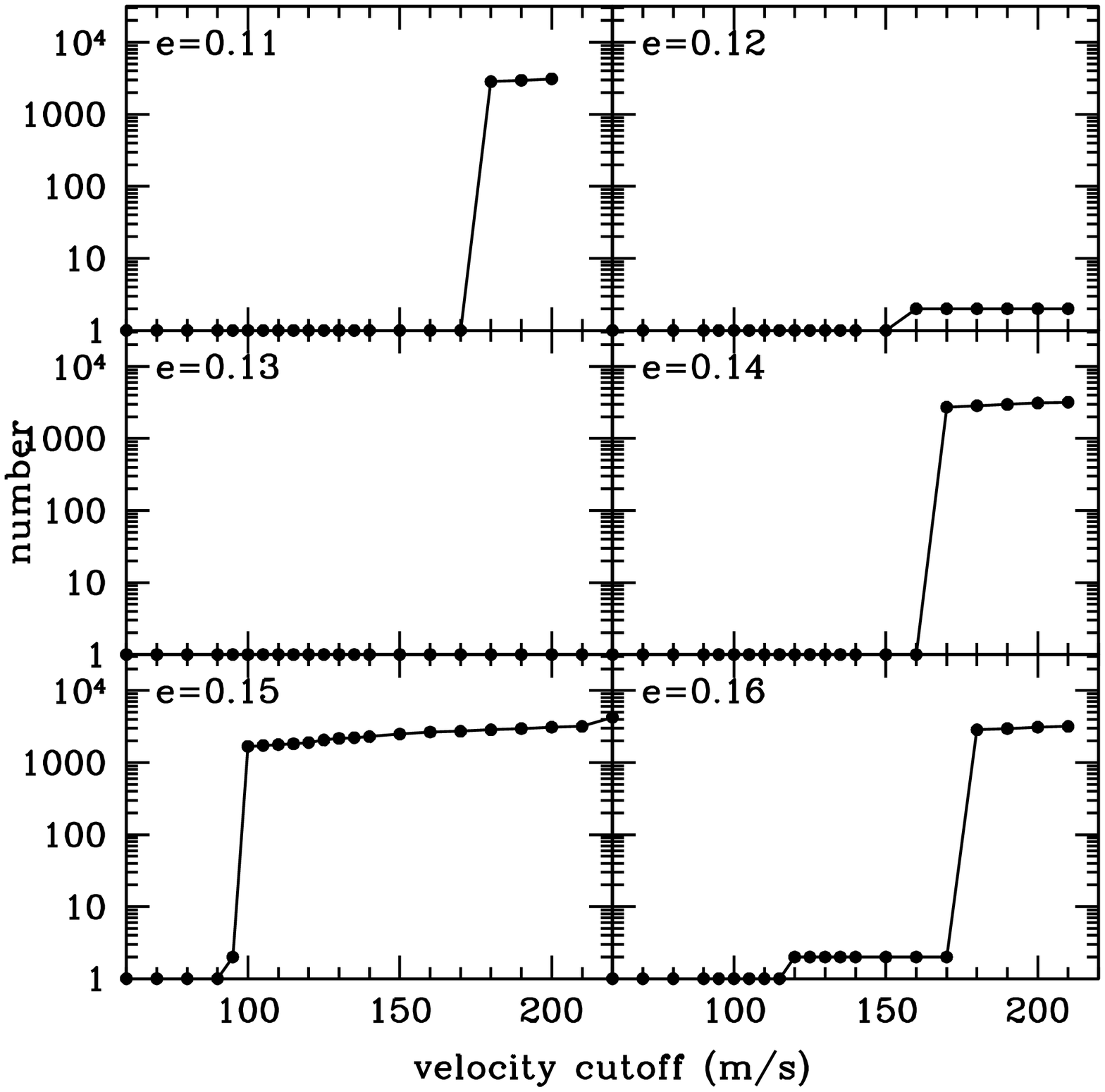}
\caption{The number of family members found as a function of velocity
  cutoff, $V_c$ (m/s), using the HCM family search algorithm. Each panel is a
  result using a different proper eccentricity for (495) Eulalia. The
  results for $e$=0.15 show a jump to large numbers of family members
  at the lowest velocity cutoff ($V_c < 100$ m/s), and the slow upward
  trend after that is typical of HCM analysis of asteroid families. }
\label{fig:HCMall}
\end{figure}

The HCM results for $e$=0.14 and $e$=0.15 show the traits of the
typical asteroid family behavior, with a sharp jump in asteroid family
membership numbers at a specific $V_c$, followed by a slow increase
thereafter (Fig. \ref{fig:HCMall}). This is not surprising, as the
low-albedo component of the Nysa-Polana complex has a mean
eccentricity of $e$=0.152. Neither case, however, shows the
characteristic jump in family membership number after exceeding the
critical velocity cutoff for the family, as seen in many examples of
Vokrouhlick\'{y} et al. (2006a). It is more similar to the results for the large
and old family Eos (Vokrouhlick\'{y} et al. 2006b). The $e$=0.15 case has
the lowest $V_c$ jump in family numbers, at $V_c=100$~m/s, and thus $e$=0.15 is 
the favored case.

\begin{figure}[h!]
\includegraphics[angle=0,width=\linewidth]{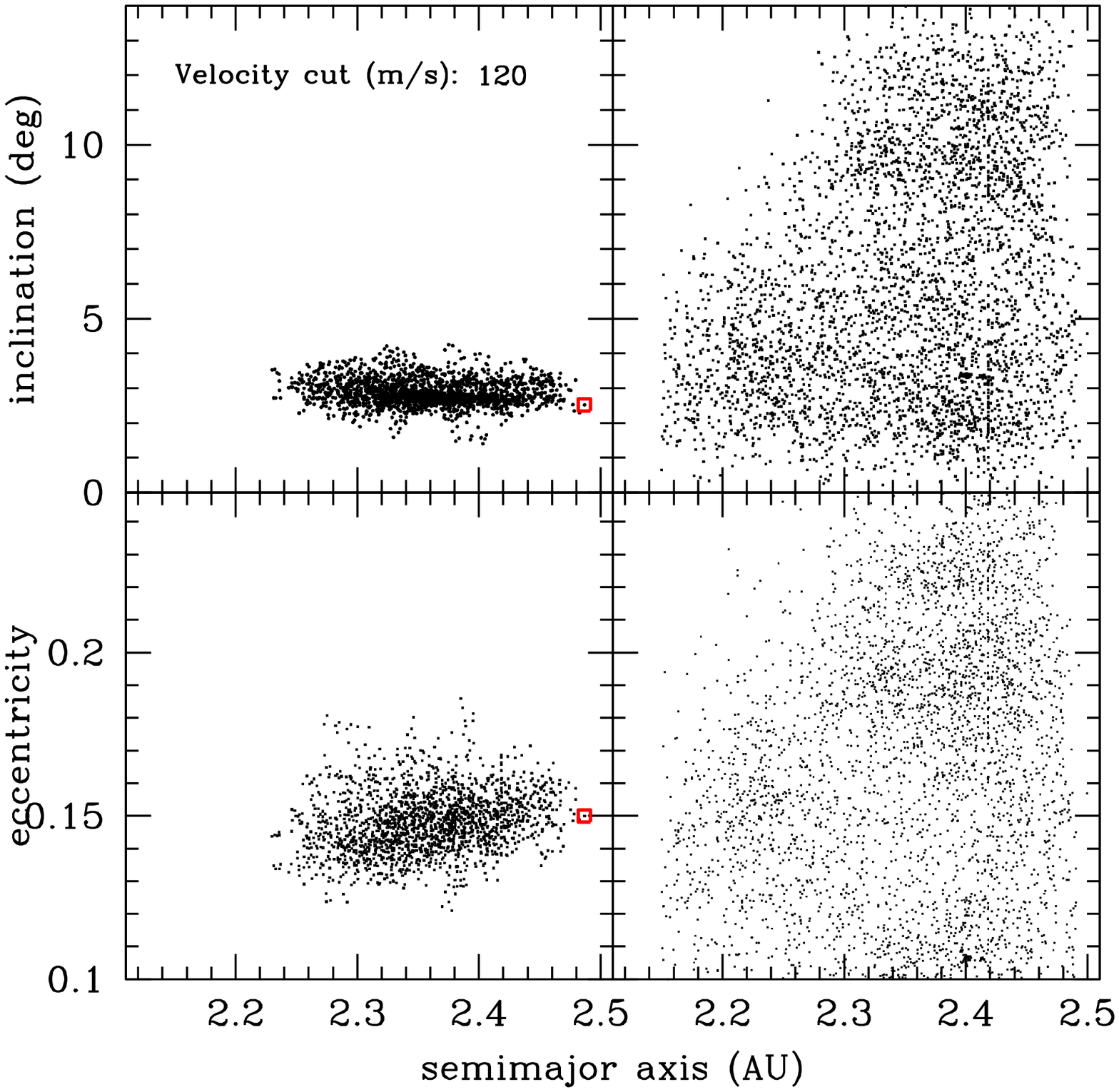}\\
\caption{The Eulalia family members plotted as their proper
  inclination (top) and proper eccentricity (bottom) as a function of
  their semimajor axis (AU). The left plots show the selected family
  members and the right side shows the asteroids in the sample not
  selected as part of the family. This plot is done for
  V$_{c}$=120~m/s, where Eulalia had an orbital eccentricity of
  $e$=0.15, and the location of (495) is shown with a red square. Note that other large families, such as Erigone have been
  removed prior to the HCM. \label{fig:EuHCM}}
\end{figure}

\begin{figure}[h!]
\includegraphics[angle=0,width=\linewidth]{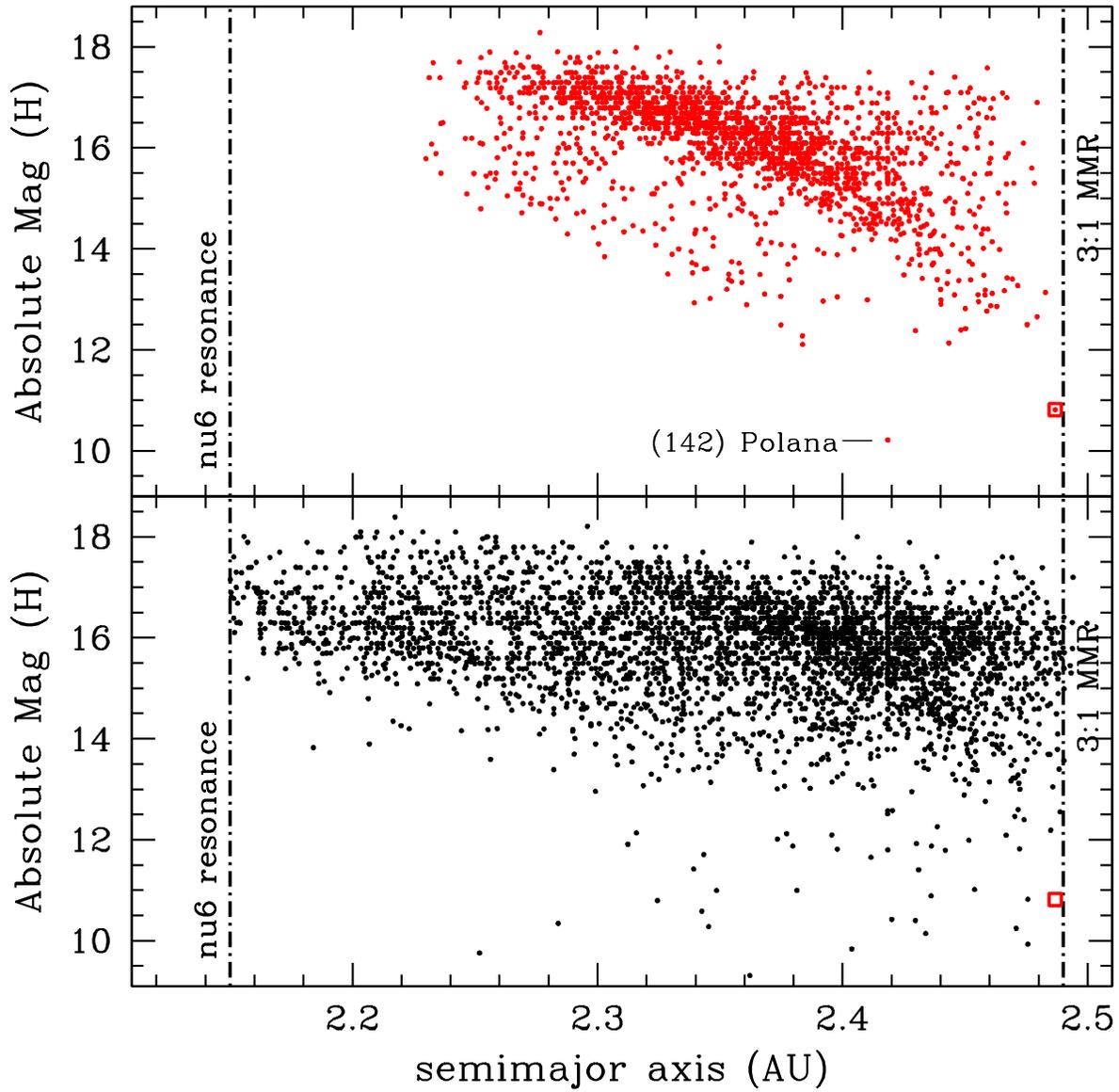}\\
\caption{The Eulalia family members plotted as their absolute
  magnitude $H$ as a function of their semimajor axis (AU). This plot is
  done for a V$_{c}$=120~m/s, where Eulalia had an orbital
  eccentricity of $e$=0.15 and its location is shown with a red square. The top panel are the members found in the
  HCM search, and the bottom panel shows the asteroids in the sample
  not selected as part of the family. Note that other large families, such as Erigone have been
  removed prior to the HCM.}
\label{fig:EuAH}
\end{figure}

One outcome from the initial HCM results is the inclusion of
asteroid (142) Polana in the family at $V_{c}=100$~m/s
(Figs. \ref{fig:EuHCM} and \ref{fig:EuAH}). It seems very clear from the
results in Fig. \ref{fig:Vshape} that multiple families are
overlapping and being selected by the HCM routine. This highlights the
problems with differentiating membership between the two families for
the bodies within the Yarkovsky envelopes of each. The inclusion of
this secondary family will pose problems in later sections when the
bounding envelope of the ``V-shape'' distribution is used to calculate
the age of the family. In fact, by estimating the lower bound of the
family in $H$ vs. $a$ space by the Yarkovsky drift curves, we can
eliminate numerous interlopers that belong to another family, rather
than the low-albedo part of the Nysa-Polana complex.

\begin{figure}[h!]
\includegraphics[angle=0,width=\linewidth]{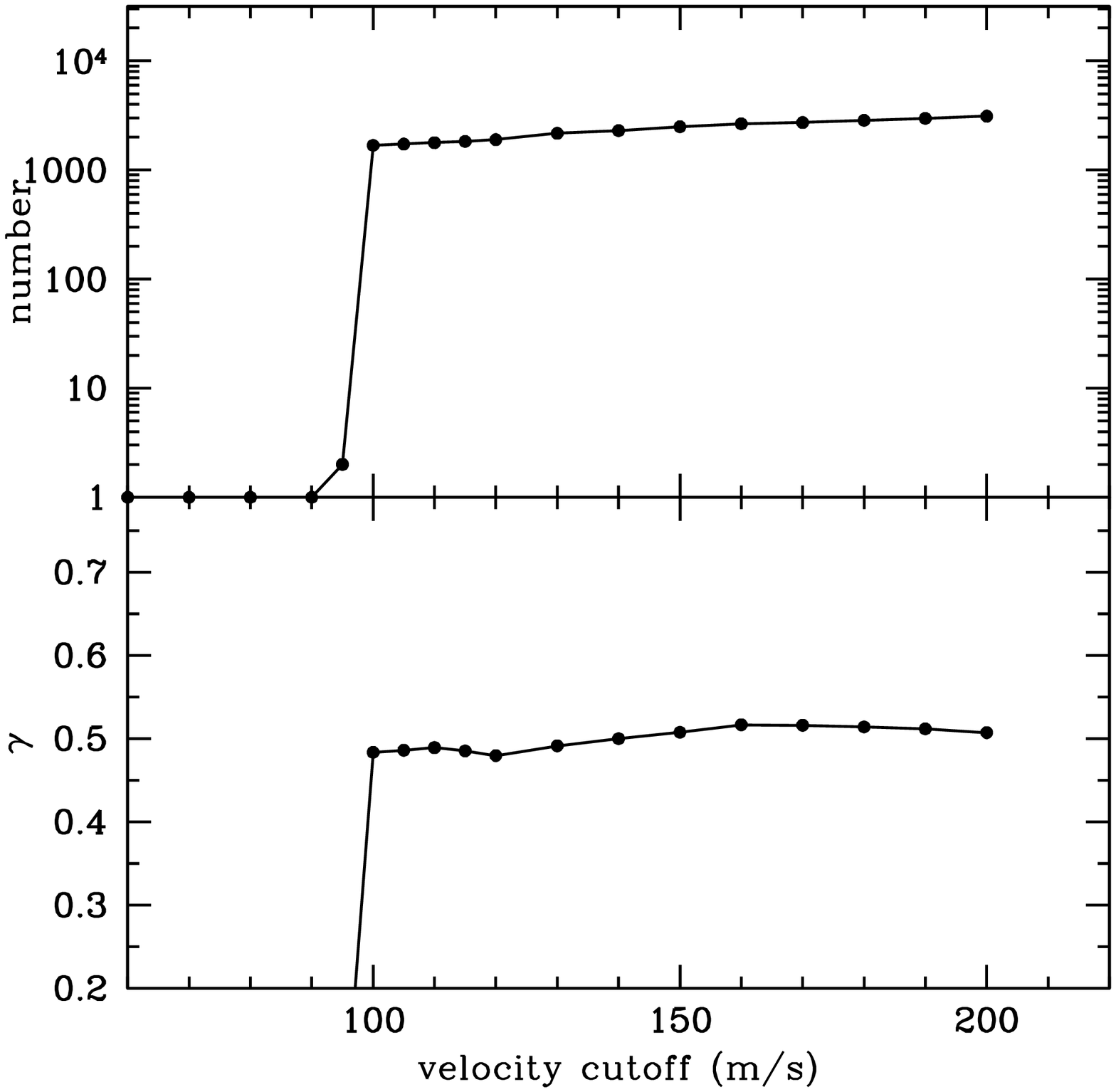}
\caption{The top panel shows number of family members found as a
  function of velocity cutoff ($V_c$) using the HCM family search
  algorithm for the case where the proper eccentricity of Eulalia is
  $e=0.15$. The bottom panel shows the exponent $\gamma$ for the power
  law size frequency distribution for $13.5 < H < 15.5$. There is no
  discernable change in the size frequency distribution for different
  velocity cutoffs as is seen for younger families.  }
\label{fig:SFDvel}
\end{figure}

The slope of the cumulative distribution of absolute magnitude $H$
value for the HCM-determined family members is sometimes used to help
select viable ranges of $V_c$ from an HCM search. This calculation was
performed on the selected family at each increment in $V_c$ (see
Fig. \ref{fig:SFDvel}). This calculation follows Vokrouhlick\'{y} et
al. (2006a), by using the power-law of the form $N(< H) \sim
10^{\gamma H}$ where $13.5 < H < 15.5$. The results show no clear
trend, and rather a low $\gamma \sim 0.5$ compared to what is found
for some young ($\sim$100s~Myr) families such as Erigone $\gamma \sim
0.8$ (Vokrouhlick\'{y} et al. 2006a) and is closer to the entire
asteroid belt population with $\gamma \sim 0.61$ (Ivezi\'{c} et
al. 2001).  Young families can have a range of size frequency
distributions with slopes sometimes dramatically different than the
background population. Over time collisional evolution will change the
size frequency distribution eventually driving it to similar values as
the background (Morbidelli et al. 2002; Bottke et al. 2005). In this
case, where there is no distinction in the size frequency distribution
from the background, this property cannot be used to either secure its
selection as a family or to make a statement about its age.

We adopt asteroid (495) Eulalia as the parent of the low-albedo
component of the Nysa-Polana complex - now referred to as the
``Eulalia family''. Its proper semimajor axis at $a=2.488$~AU is
adopted as the center of the family. In the next section we can
determine the extent and age of the family by using the members found
in this HCM clustering search, and we will use the family as defined
by the HCM criteria with $V_c=120$~m/s, which by visual inspection was preferable to the $V_c=100$~m/s selection due to larger numbers without increasing the size of the family in $a$, $e$, or $i$.

\subsection{Age estimate for the Eulalia family}
\label{s:Eage}

The age of the family can be estimated by the spreading in semimajor
axis of family members over time, which is a function of an asteroid's
size, thermal properties and spin axis (Bottke et al. 2006). The size
or absolute magnitude of many main belt asteroids is well measured
through the combination of recent WISE results, years of optical
surveys such as Sloan Digital Sky Survey (SDSS) and past infrared
surveys such as IRAS and WISE (some potentially important biases in
  the absolute magnitude values are discussed in Section 4.). The thermal properties
have been directly measured for only a few asteroids (see Delb\'{o} et
al. 2007), but have been inferred for populations due to ensemble
behavior for asteroid families with a range of ages (Bottke et
al. 2006). The evolution of the spin axis of a body is related to its
thermal properties and the thermal YORP-effect, and this behavior,
though still being investigated, is constrained by asteroid family
investigations (Vokrouhlick\'{y} et al. 2006a; Bottke et
al. 2006). Similarly, the initial distribution of fragment orbits
following an asteroid impact provides an uncertain initial condition
preceeding the family spreading.

Vokrouhlick\'{y} et al. (2006a) have previously employed a complex
fitting routine that simultaneously solves for family age, YORP
reorientation timescales and strength, and a size-dependent initial
velocity dispersion. Given the large extent of the Eulalia family,
that at least half of the family is lost into the 3:1~MMR, and its
location overlapping another primitive family, we focus here on the
family boundaries in order to estimate family age. A more detailed
method of the age determination, solving for YORP reorientation and
the initial velocity disperion, will be presented in later works.

The $C$ distribution of asteroids was used to locate the center of the
family, based on a cluster of similar values. A bounding $C$ value for
the family can be used to estimate an age for the family, based on the
maximum Yarkovsky drift timescales.  Using the $C$ value we find,
\begin{equation}
0.2H=\log(\Delta a/C)
\end{equation}
where $\Delta a=(da/dt)T$, and
$(da/dt)=(da/dt)_0(D_0/D)$ with $(da/dt)_0$ being the maximum Yarkovsky
drift rate for a diameter $D_0$ ignoring any spin axis variation. It will be
useful to use $D=D_0\times10^{-H/5}/\sqrt{\pV}$, and combine them so
that

\begin{equation}
C=\left(\frac{da}{dt}\right)_0\sqrt{\pV}\; T
\end{equation}

\noindent where $(da/dt)$ is $\sim 3\times 10^{-5}$~AU/Myr for a 5~km
asteroid in the inner Main Belt with a density of 2.5~\gcc and thermal
conductivity $K \sim 0.01-0.001$~W~m$^{-1}$~K$^{-1}$ (taken from
Bottke et al. 2006). The formulation uses a $D_0=1329$~km, so the
nominal drift rate (in AU/Myr) is,
\begin{equation}
\left(\frac{da}{dt}\right)_0=3.0\times 10^{-5}\left(\frac{2.5~\mathrm{g~cm^{-3}}}{\rho}\right)\left(\frac{5~\mathrm{km}}{D_0}\right)\left(\frac{\mathrm{AU}}{\mathrm{Myr}}\right).
\end{equation}
For an $H=0$ with $D_0=1329$~km and a density of 1~\gcc~the rate is
$\left(\frac{da}{dt}\right)_0=2.8\times 10^{-7}$~AU/Myr. There is no
published measurement of the density of any family members, so we
selected our nominal value of 1~\gcc\ for simplicity and the few
measurements of C-type asteroids clustering around this value; with
(379) Huenna at $\rho\sim0.9$~\gcc\ (Marchis et al. 2008), and (253)
Mathilde with $\rho\sim1.3$~\gcc\ (Yeomans et al. 1997) are two with
accurate estimates. The drift rate is inversely proportional to
the density, and thus this estimate can be a systematic source of
error in our calculations. Now we can estimate the age based on the
estimates for $C$ by,
\begin{equation}
T=C/\left(\sqrt{\pV}\left(\frac{da}{dt}\right)_0\right)
\end{equation}
with an albedo $\sim$~0.052 for (495) Eulalia.

\begin{figure}[h!]
\includegraphics[angle=0,width=\linewidth]{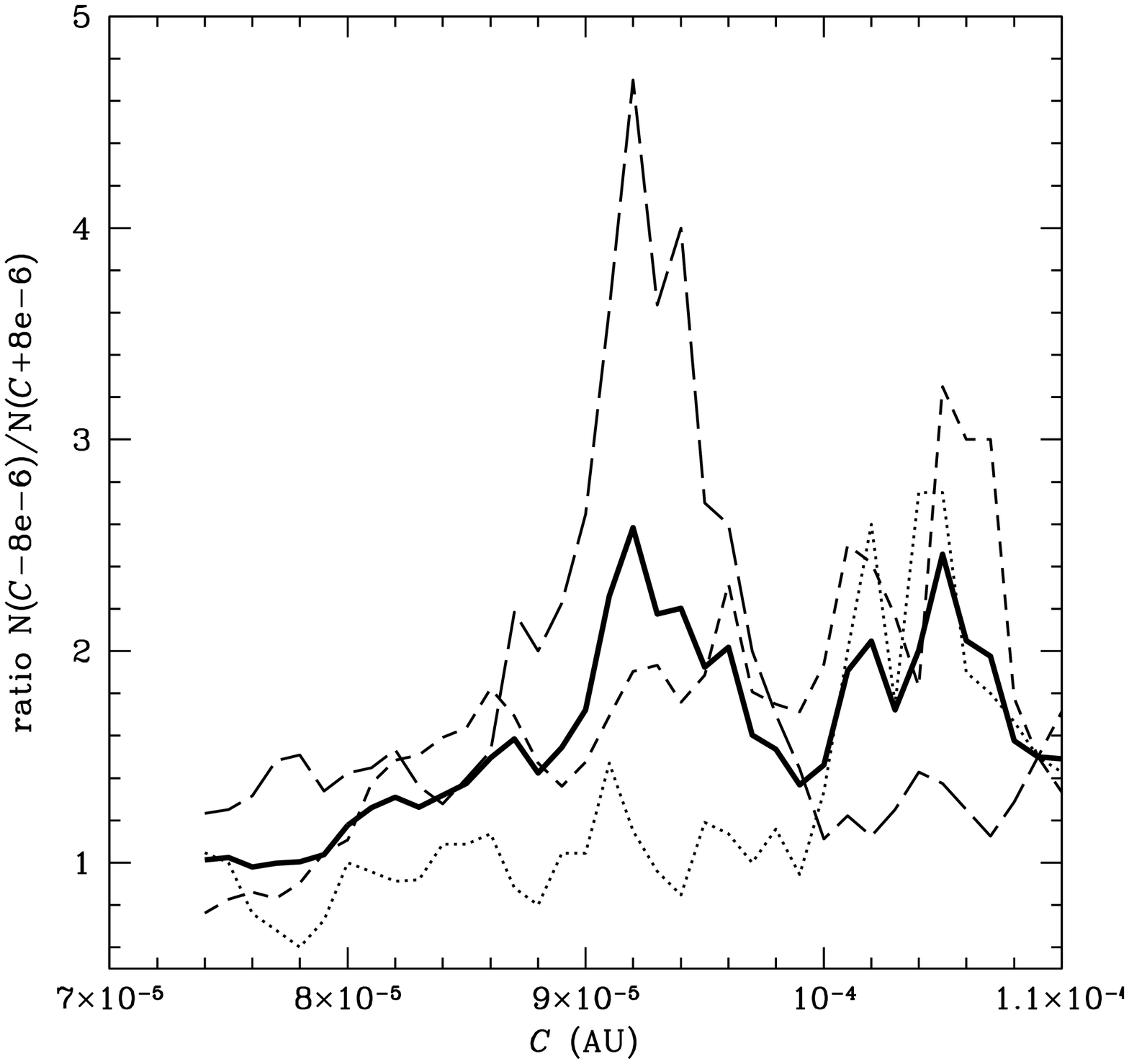}
\caption{ The ratio of the asteroids with ($C-8\times10^{-6}$)~AU over
  the number with ($C+8\times10^{-6}$)~AU, where a large value represents
  a strong density contrast at the tested value of $C$. This test was
  done for the family center at $a_{\mathrm{init}}=2.488$~AU, and combined ratio values
  for three different ranges of absolute magnitude $H$: the dotted
  line is $13.5 < H < 15$, the short dashed line is for $15 < H < 16$,
  and the long dashed line is for $16 < H < 16.5$. The dark solid line
  is the average of the three, with the value of $C=9.2\times10^{-6}$~AU
  is slightly preferred over the value of $C=10.5\times10^{-6}$~AU due to
  the strong preference at smaller asteroids ($16 < H < 16.5$) for the
  former fit.
\label{fig:AgeFit}}
\end{figure}

To calculate the age of the family we are interested in the
boundary of the $C$ distribution, or the envelope of Yarkovsky lines
for $H$ as a function of $a$. We devised a fitting routine whereby the
value of $C$ was varied and the ratio of asteroids with
$C-8\times10^{-6}$~AU was compared with $C+8\times10^{-6}$~AU. A strong
contrast in numbers indicates the boundary of the family has been
reached. This was done for all the members selected by the HCM routine
for $V_{c}=120$~m/s in steps of $\Delta C=1\times10^{-6}$~AU (where these bin sizes were selected via trial and error testing). Due to
the increasing number of asteroids at greater $H$ (smaller sizes), we
measured this ratio for three different size ranges, $13.5 < H_i < 15
< H_{ii} < 16 < H_{iii} < 16.5$. The average of the three ratios finds
two peaks, at $C=9.2\times10^{-5}$~AU and $C=10.5\times10^{-5}$~AU
(Fig. \ref{fig:AgeFit}). The smaller value of $C=9.2\times10^{-5}$~AU is
favored by the smaller asteroids, while $C=10.5\times10^{-5}$~AU is
favored by the larger asteroids. This could be a real effect caused by
thermal effects acting differently at small sizes, or simply confusion caused by interlopers.

\begin{figure}[h!]
\includegraphics[angle=0,width=\linewidth]{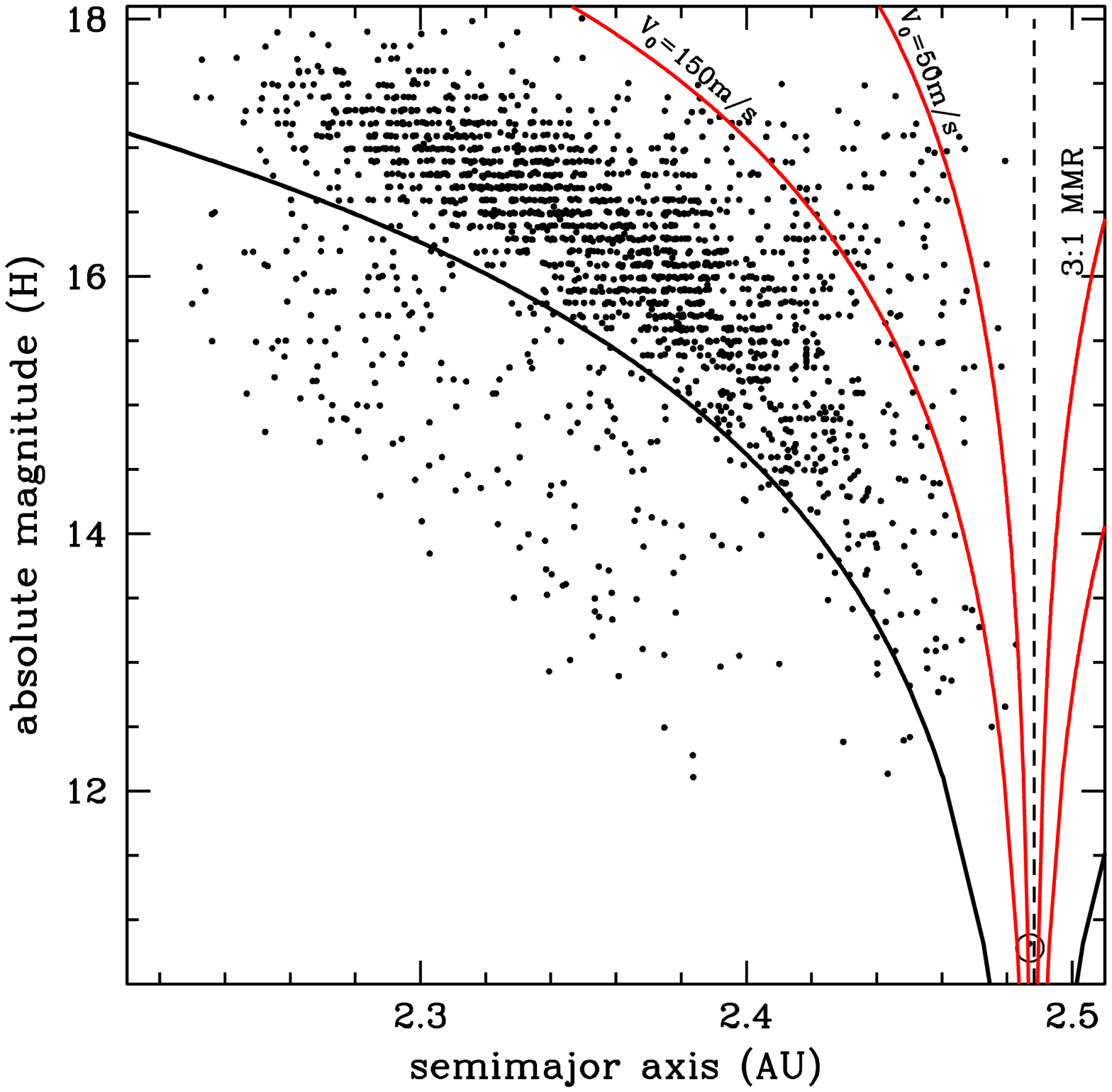}\\
\caption{Best fit boundary for the Eulalia family. The Yarkovsky line
  for the family boundary at $C=10.5\times10^{-5}$~AU is the solid black
  line.  The red lines represents a simple initial dispersion for a
  velocity dispersion of fragments for $V_0=50$~m/s and $V_0=150$~m/s by
  $\Delta a=(2/n)V_0(5~\mathrm{km}/D)$ where $n$ is the mean
  motion. \label{fig:TestAge10.5}}
\end{figure}

\begin{figure}[h!]
\includegraphics[angle=0,width=\linewidth]{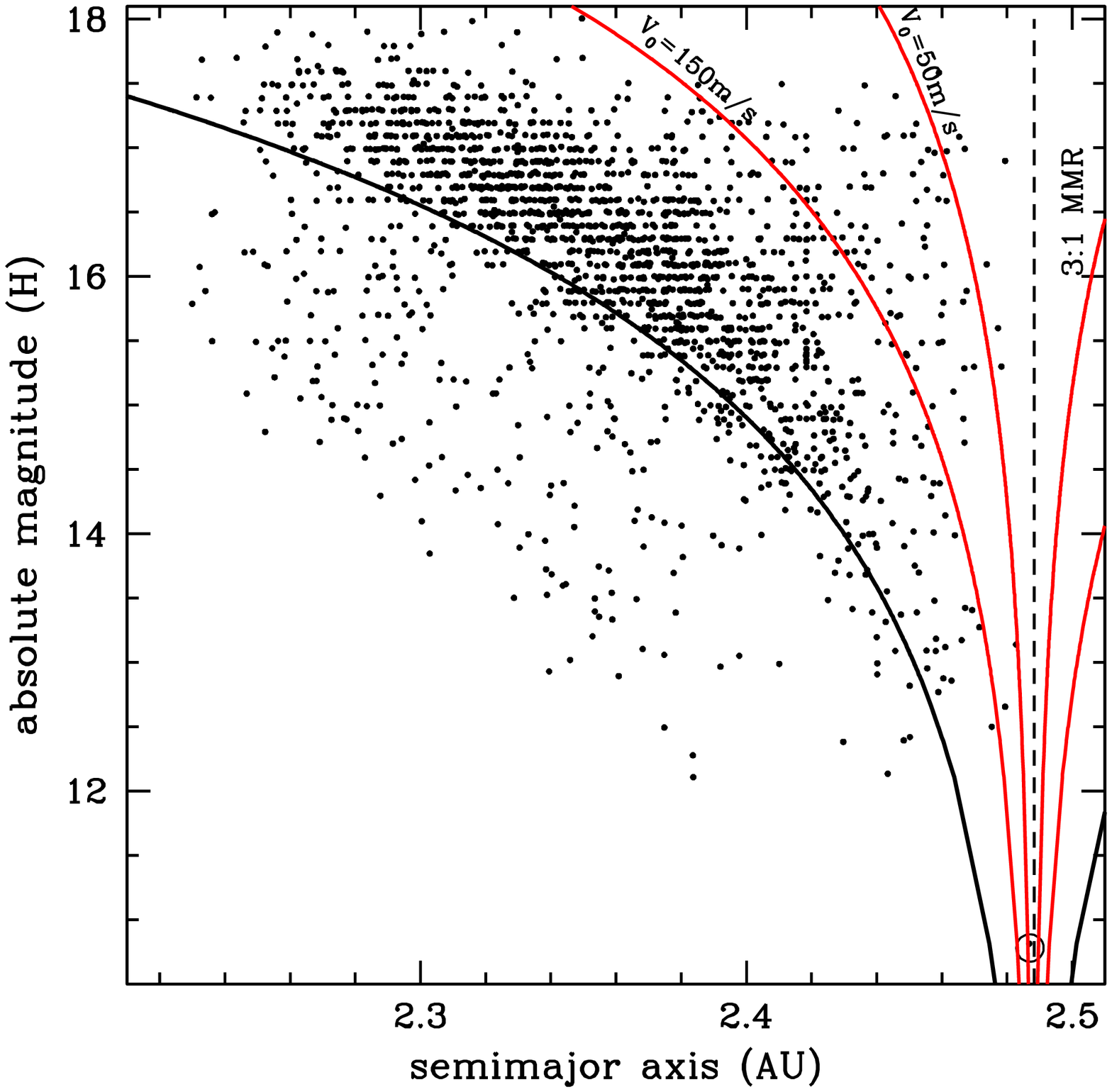}\\
\caption{Best fit boundary for the Eulalia family. The Yarkovsky line
  for the family boundary at $C=9.2\times10^{-5}$~AU is the solid black
  line.  The red lines represents a simple initial dispersion for a
  velocity dispersion of fragments for $V_0=50$~m/s and $V_0=150$~m/s by
  $\Delta a=(2/n)V_0(5~\mathrm{km}/D)$ where $n$ is the mean
  motion. \label{fig:TestAge9.2}}
\end{figure}

The best fit boundaries are plotted in Figs. \ref{fig:TestAge10.5} and
\ref{fig:TestAge9.2}.  Alone, these values of $C$ would correspond to
ages of $T\sim1440$ and $T\sim1644$~Myr. However, the calculation does
not yet account for the initial fragment semimajor axis distribution
produced immediately after the family forming event.

The size-dependent velocity dispersion of the fragments due to the
asteroid impact creates a distribution of $a_{init}$  preceeding any Yarkovsky drift. Again,
following Vokrouhlick\'{y} et al. (2006a), we can estimate
size-dependent $a_{init}$  for collision outcomes with different
velocity dispersions, where here we calculate another $\Delta a$ as the distance from the center of the family,
\begin{equation}
\Delta a = \frac{2}{n}V_\mathrm{T} + {\cal O}(e)
\end{equation}
where $n$ is the mean motion of the target asteroid and $V_\mathrm{T}$
is the transverse velocity of the fragment and ${\cal O}(e)$
  simply denotes other eccentricity-dependent terms. We can use a
very idealized equation,
\begin{equation}
V_\mathrm{T}=V_0\left(\frac{D_0}{D}\right)
\end{equation}
Vokrouhlick\'{y} et al. (2006a) uses $V_{SD}= V_0(5~\mathrm{km}/D)$, where
$V_{SD}$ is the standard deviation of a directional component of a
velocity field following an asteroid disruption. The $V_{SD}$ can be
equated to $V_\mathrm{T}$ in this simple estimation, and $V_0$ can be
equated very roughly to the escape speed at the surface of the target
body, which scales linearly with diameter. So that,
\begin{equation}
\Delta a = \frac{2}{n}V_{\mathrm{T}} + {\cal O}(e) \sim\frac{2}{n}V_{0}\left(\frac{5~\mathrm{km}}{D}\right),
\end{equation}
and using a mean motion for the target body, Eulalia in this case, the $\Delta a$ can be calculated as a function of fragment size.

We have not yet estimated the total mass in the family or
estimated the size of the parent of the family. This calculation
is made later in Section \ref{s:agesSFD}, finding a parent body
$D=100-160$~km. That is used as a guide here since the velocity
  dispersion should be similar to the parent body escape speed that
  scales linearly with the parent body size. So we include
calculations of the estimated size-dependent $\Delta a$ due to the
initial collision event with $V_{0}$=50~m/s and $V_{0}$=150~m/s (see
the red lines in Figs. \ref{fig:TestAge10.5} and
\ref{fig:TestAge9.2}). Since the $\Delta a$ from the initial fragment
displacement follows the same $1/D$ relationship as the Yarkovsky
drift, we can relate the initial fragment distribution in terms of
$C$, and simply subtract off this $C$ and re-calculate the age from
the total Yarkovsky drift.  These distributions correspond to a
maximum initial semimajor axis $C=1.13\times10^{-5}$~AU and
$3.39\times10^{-5}$~AU, such that the best-fit bounding value of
$C=9.2\times10^{-5}$~AU, would simply be a $C=8.07\times10^{-5}$~AU
and $5.81\times10^{-5}$~AU, which correspond to ages of $T=1264$ and
$910$~Myr. While for $C=10.5\times10^{-5}$~AU, would correspond to
total ages of $T=1468$ and $1114$~Myr.

The relationship between these calculations are shown in
Fig. \ref{fig:TestAge9.2}, where the black line is the $\Delta a$ due to
Yarkovsky drift from a center at $a=2.488$~AU with an age of 1160~Myr. The red
lines are the $\Delta a$ due to the fragment velocity dispersion of
50 and 150~m/s.  From these calculations we find the age of the
family between 900--1500~Myr.

One important caveat in these simplistic age calculations is the role
of obliquity changes due to the Yarkovsky-O'Keefe-Radzievskii-Paddack
(YORP) effect (Rubincam 2000; Vokrouhlick\'{y} and \v{C}apek 2002;
Bottke et al. 2002). The YORP effect acts by the asymmetrical
reflection and re-emission of thermal radiation to affect the rotation
state of an irregularly shaped body. When the YORP effect changes the
obliquity of an asteroid, the semimajor axis drift rate due to the
Yarkovsky effect will also change. The preferred spin states are
those that maximize the Yarokovsky drift (0$^\circ$ and 180$^\circ$),
such that family members with initial obliquities resulting in slow
Yarkovsky drift are expected to evolve over time to a faster-drifting
state. Large uncertainty exists in modeling the evolution of the YORP
re-orientation of obliquities due to the YORP evolution of the
rotation rate as well and complications at very high and very low spin
rates. An estimate of re-orientation timescales is $\sim$~300--600~Myr
for $\sim$~5~km objects (Vokrouhlick\'{y} and \v{C}apek 2002;
\v{C}apek and Vokrouhlick\'{y} 2004).  Objects well above this size
are then likely to have not experienced a full YORP reorientation
since the formation of the family. Any dramatic effects due to YORP
cycles for this family would therefore not be significant for
asteroids larger than $H < 14$ to be conservative.

\begin{figure}[h!]
\includegraphics[angle=0,width=\linewidth]{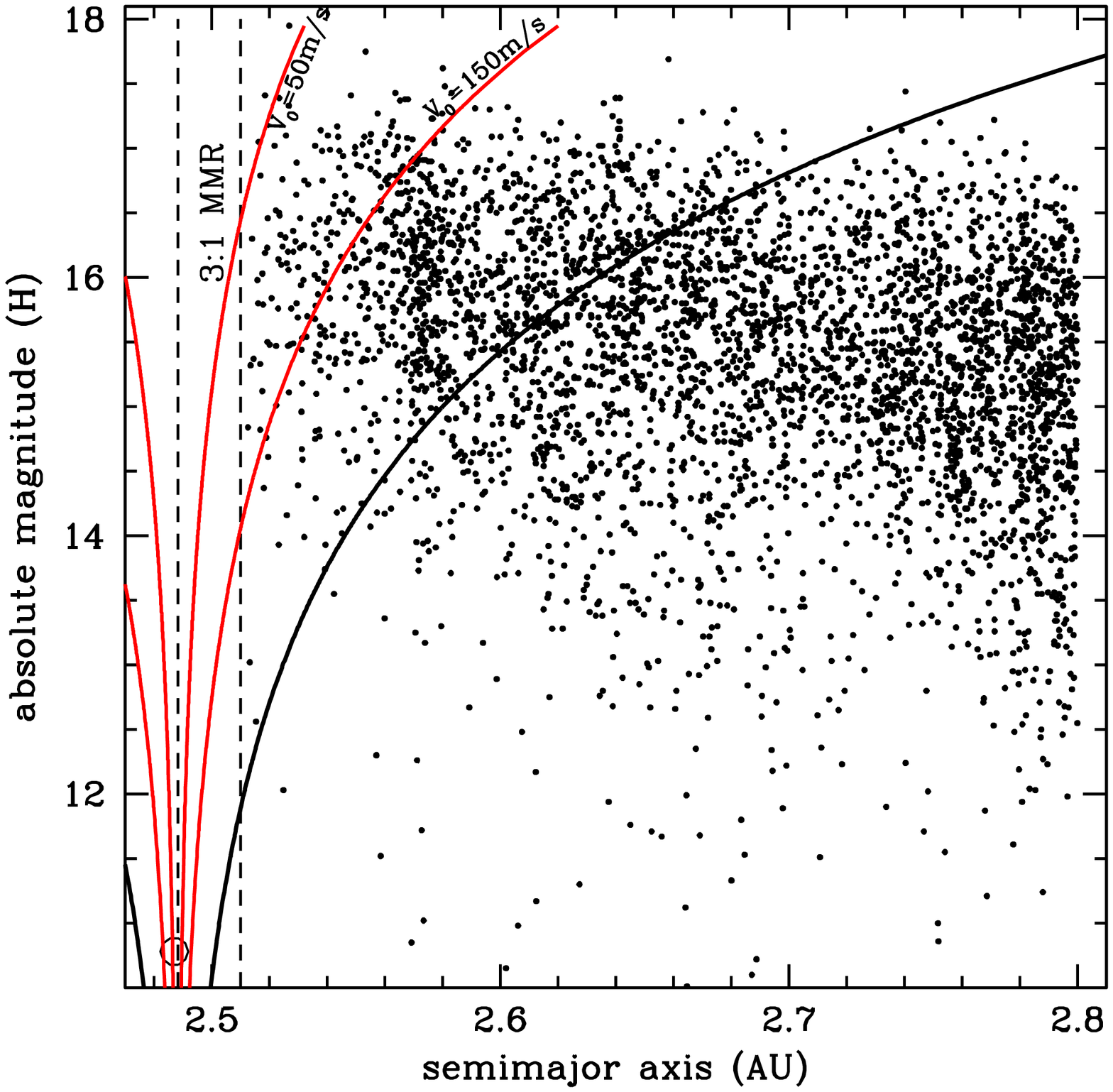}\\
\caption{Asteroids measured by WISE to have \pV$ < $0.1 with $0.1 < e <
  0.2$ and inclination $< 10^\circ$ with semimajor axis $a >
  2.5$~AU. The Yarkovsky line for the family boundary at
  $C=9.2\times10^{-5}$~AU is the solid black line.  The red lines
  represents a simple initial dispersion for a velocity dispersion of
  fragments for $V_0=50$~m/s and $V_0=150$~m/s. This selection of
  asteroids, matching the orbital properties of the detected family
  but on the other side of the 3:1~MMR show no sign of a family
  signature. \label{fig:Eul31}}
\end{figure}

One possible tracer of $V_{0}$ would be family members reaching
the {\it other} side of the 3:1 resonance (for $a > 2.5$~AU). 
Given
the center of the family and the Yarkovsky drift boundaries from the
asteroids with $a<2.5$~AU, we selected the subset of asteroids with
$a>2.5$~AU that would host any family members.
There is no
clear evidence of a high density of asteroids around the Yarkovsky
lines (Fig. \ref{fig:Eul31}), confirmed by the distribution of $C$
values around an $a_{\mathrm{init}}=2.487$~AU. The lack of existence of,
or inability to detect, members on the far side of the 3:1 MMR limits
the $V_{0}$ for the fragment distribution. Very large velocity
distributions would place many asteroids on orbits beyond the 3:1~MMR,
where they could be observable today.  For a conservative boundary of
the 3:1~MMR of $a\sim2.51$~AU at the relevant eccentricities of $0.1 < e < 0.2$
a $V_{0}=50$~m/s would only place some objects of $H>15$ beyond the
resonance, while at $V_{0}=150$~m/s only some with $H>13.5$. While a
substantial mass of the family can be contained in bodies of these
sizes and smaller, it also implies a large fraction of the mass never
made it past the 3:1~MMR. Thus the $V_{0}$ estimates used in this
work, presented to bracket possible ranges of fragment velocity
dispersion, also are consistent with the lack of a clear family signature
across the 3:1~MMR.

\section{Properties of the Eulalia family members}
\label{sec:Properties}

\subsection{The SFD and total mass of the Eulalia family}
\label{s:agesSFD}

The size frequency distribution of the family can be estimated
from the objects selected in the HCM routine, using the power-law of
the form $N(< H) \sim 10^{\gamma H}$. In Section 2.4 a $\gamma$ for only the
range of $13.5 < H < 15.5$ was compared to the background, now we
consider the size distribution of all prospective family members,
using it to estimate the mass of the entire family.

Simply summing the mass of all observed bodies is a quick estimate of
the original parent's size. However, we are only now detecting half of
the family, as the other half were launched into, or very quickly
entered, the 3:1~MMR and were lost. There is another factor of two, as
only bodies launched to smaller semimajor axis survive the initial
displacement from the parent, but then any asteroid with an initial
outward Yarkovsky drift will very quickly drift into the 3:1~MMR.
Similarly, the dataset is missing objects as the WISE survey is not
complete, even at large sizes. A comparison of the WISE survey and
databse of known asteroids, for large asteroids in the inner Main Belt,
found that it detected $\sim$80\% of all asteroids with $H<10$, with
that fraction dropping to 55\% for $H<15$. Combined with the previous
family scaling factor of 4, scaling by 5 to account for the $\sim$80\%
WISE detection efficiency is closer, though this scaling factor should
grow at smaller sizes.  Finally, the numbers are scaled at each size
according to the WISE detection efficiency. It should be noted that
collisional evolution could play a major role in altering the family
SFD over time, and this should be considered in future models.

\begin{figure}[h!]
  \includegraphics[angle=-90,width=\linewidth]{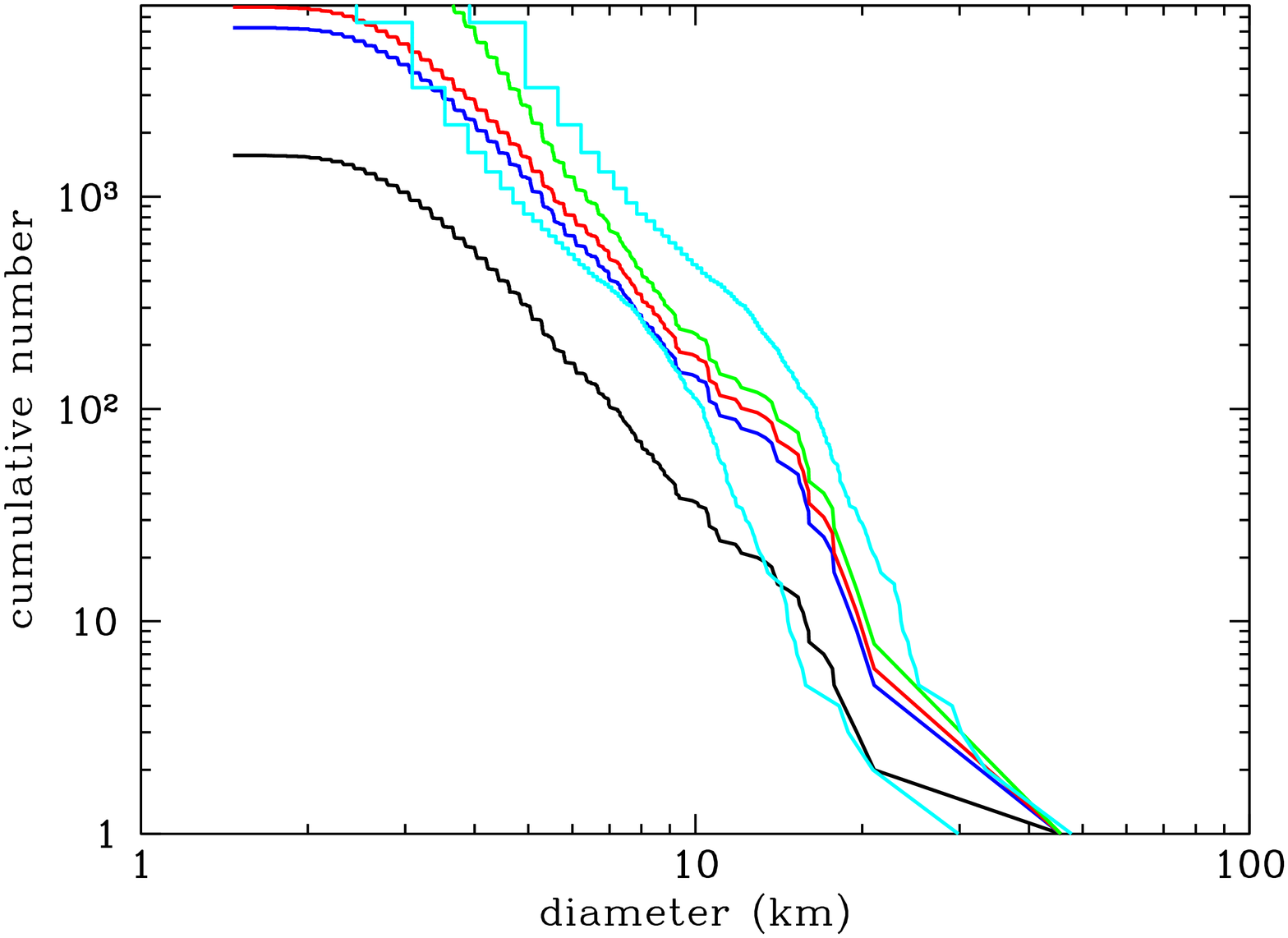}
  \caption{Calculated size frequency distribution of the Eulalia
    family for an HCM with $V_{c}=120$~m/s and
    $C=9.2\times10^{-5}$~AU. The cumulative number of fragments with a
    size $D$ is plotted as a function of the asteroid diameter. The
    black line is taken directly from the HCM with
    $V_{c}=120$~m/s, while the blue line increases the
    family for all members other than the parent (495) Eulalia by a
    factor of four. The red line is the family scaled up by a factor
    of 5. The integrated size for the black line is $D=68.6$~km and
    for the blue is $D=100.1$~km, and for the red line is
    $D=107.2$~km.  The green line is the family size accounting for
    the factor of 4, and then increased according to the WISE
    detection efficiency at each size. The integrated size of that
    curve is $D=121.2$~km.  The cyan lines are the result from
    Benavidez et al. (2012) for 100~km basalt rubble pile target with
    a 3~km/s 34~km impactor hitting at a 15$^\circ$ angle. The largest
    remnant in the simulation outcome is $D=29.8$~km, slightly smaller
    than (495) Eulalia ($D=40$~km).
 \label{fig:SFDtest}
}
\end{figure}

The integrated size for the selected family members from the
$V_\mathrm{c}=120$~m/s HCM routine is $D=68.6$~km (the black line
in Fig. \ref{fig:SFDtest}). Note that this includes a limiting $H$ vs. $a$
criteria to eliminate interlopers. When a factor of 4 for the loss of
3/4 of the family to the 3:1~MMR is accounted for (except the largest
body, parent Eulalia) the estimate increases to $D=100.1$~km (blue
line in Fig. \ref{fig:SFDtest}). Finally, a full factor of 5,
accounting for the survey limitations, equals a body with $D=107.2$~km
(red line in Fig. \ref{fig:SFDtest}). The final correction for
size-dependent WISE detection efficiency increases the size to
$D=121.2$~km (green line in Fig. \ref{fig:SFDtest}).

Since only the largest bodies are detected with very high efficiency,
an alternative method to estimate total family mass is needed. An alternative method relies on the size frequency
  distributions for asteroid families derived from Smoothed Particle
  Hydrodynamics (SPH) simulations of catastrophic asteroid impacts
  that are coupled to $N$-body gravitational simulations of their
  reaccumulation (Durda et al. 2007; Benavidez et al. 2012). This
  method matches size frequency distributions between simulations and
  observations at large fragment sizes, and then uses the simulation
  to estimate how much unobservable material may also belong to the
  family.  One SPH and $N$-Body asteroid collision simulation was
selected from the suite of simulations by Benavidez et al. (2012). It
used a 100~km basalt rubble pile target with a 3~km/s 34~km impactor
hitting at a 15$^\circ$ angle. The largest remnant in the simulation
outcome is $D=29$~km, slightly smaller than (495) Eulalia
($D=40$~km). However, the shape/slope of the curve is a reasonable
match for the Eulalia family. These simulations are based on a 100~km
target, and a match in slope must be coupled with a horizontal shift
to scale for the original parent size.  Shifting the size
  distribution by a factor of 1.6 is equivalent to increasing the
  target body size by the same factor, thereby creating a size
  frequency distribution for a $\sim$160~km target.   Given the
relative close match in both slope and horizontal scaling, this
suggests that the simple integrated size estimate of $D\sim107$~km is
close to the SPH progenitor target scaled to $\sim160$~km, and the
latter serves as a resonable upperbound given the uncertainties with
this particular family.

\subsection{Observed physical properties}\label{s:observed}

The WISE mission facilitated the analysis of this family while also
providing some physical characterization of its constituents. Other
public databases and catalogs of observations are available to assess
the physical properties of the family members. Here we summarize the
available data.

\subsubsection{Albedo distribution}

From the WISE mission, there are calculated albedos for each object in
the dataset used here. We compare the albedo distribution of the
bodies found by the HCM routine for $V_{c}=120$~m/s, with the $C=9.2\times10^{-5}$~AU boundaries.

\begin{figure}[h!]
\includegraphics[angle=0,width=\linewidth]{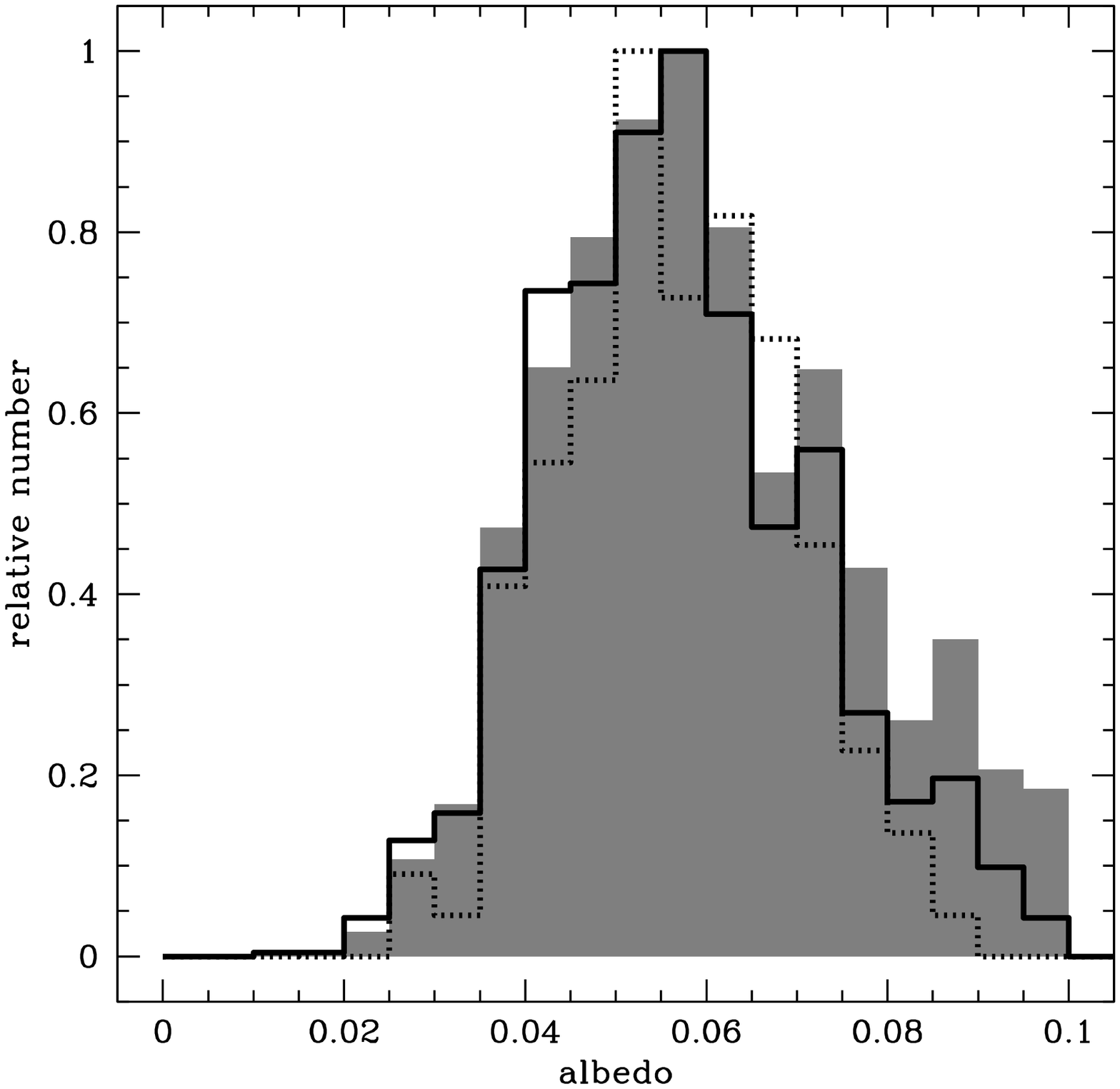}
\caption{The gray-filled histogram shows the relative distribution of
  albedos for all of the 6702 bodies selected for this study (\pv $<$ 0.1), while
  the black outline histogram shows the distribution for the selected
  Eulalia family members inside the $C=9.2\times 10^{-5}$~AU
  boundary. The dotted line is the albedo distribution for only the
  subset of the family members with $H < 15$, accounting for 128 of the 1563 total
  members plotted here. \label{fig:Albedo}}
\end{figure}

We find only a slightly tighter cluster of albedo values in this
sample compared to the background (Fig. \ref{fig:Albedo}). Given that
the entire sample is less than \pv$ < 0.1$, and typical formal errors
in albedo measured by WISE are $\sim$0.008 for $H < 15$ and
$\sim$0.018 for the entire sample, a clear cluster in albedo was
unlikely. The bulk of the asteroids included in the family are small
objects, the distribution may be dominated by objects with larger
errors in their measurement due to their small sizes. The 128
  objects with $H < 15$ show no clear distinction from the entire
  sample of 1563 selected for these boundaries of the family
(Fig. \ref{fig:Albedo}).  Pravec et al. (2012) found that there is
divergence between $H$ values found in the MPC database and those
measured through an observational campaign, where the MPC values
are systematically too small, reaching a maximum mean offset of
  -0.4 to -0.5 around $H=14$. This would lead to albedo estimates
that are slightly too large and becomes significant and
  systematic at $H$ above 10.  The WISE calculated albedos relied on
the $H$ values in the MPC database, and this therefore could account
for a source of error as well.

\subsubsection{SDSS colors}

The Sloan Digital Sky Survey (SDSS) observed 471,000 moving objects,
for which 220,000 observations were matched to 104,000 different
asteroids. The photometry was done in 5 wavelengths and is a valuable
tool for establishing taxonomic relationships for large groups of
asteroids. The five wavelengths can be combined in numerous ways in
order to capture the many possible features found in visible spectra
of different asteroid types, but a common method is to compare
the $i - z$ colors as a function of $a^\ast$ (where the $i$ filter is
centered at 769~nm, the $z$ at 925~nm and $a^\ast$ is the derived
first principal component for the distribution of asteroid colors in
the SDSS $r-i$ vs. $g-r$ color-color diagram). The former measures the
depth of the 1 micron silicate absorption bands mainly due to olivines
and pyroxenes, the latter is a proxy for the slope of the spectra (see
Fig 3. in Parker et al. 2008). In this space, primitive asteroids of
the C- and B-types are typically found with negative slope ($a^\ast <
0$).

\begin{figure}[h!]
\includegraphics[angle=0,width=\linewidth]{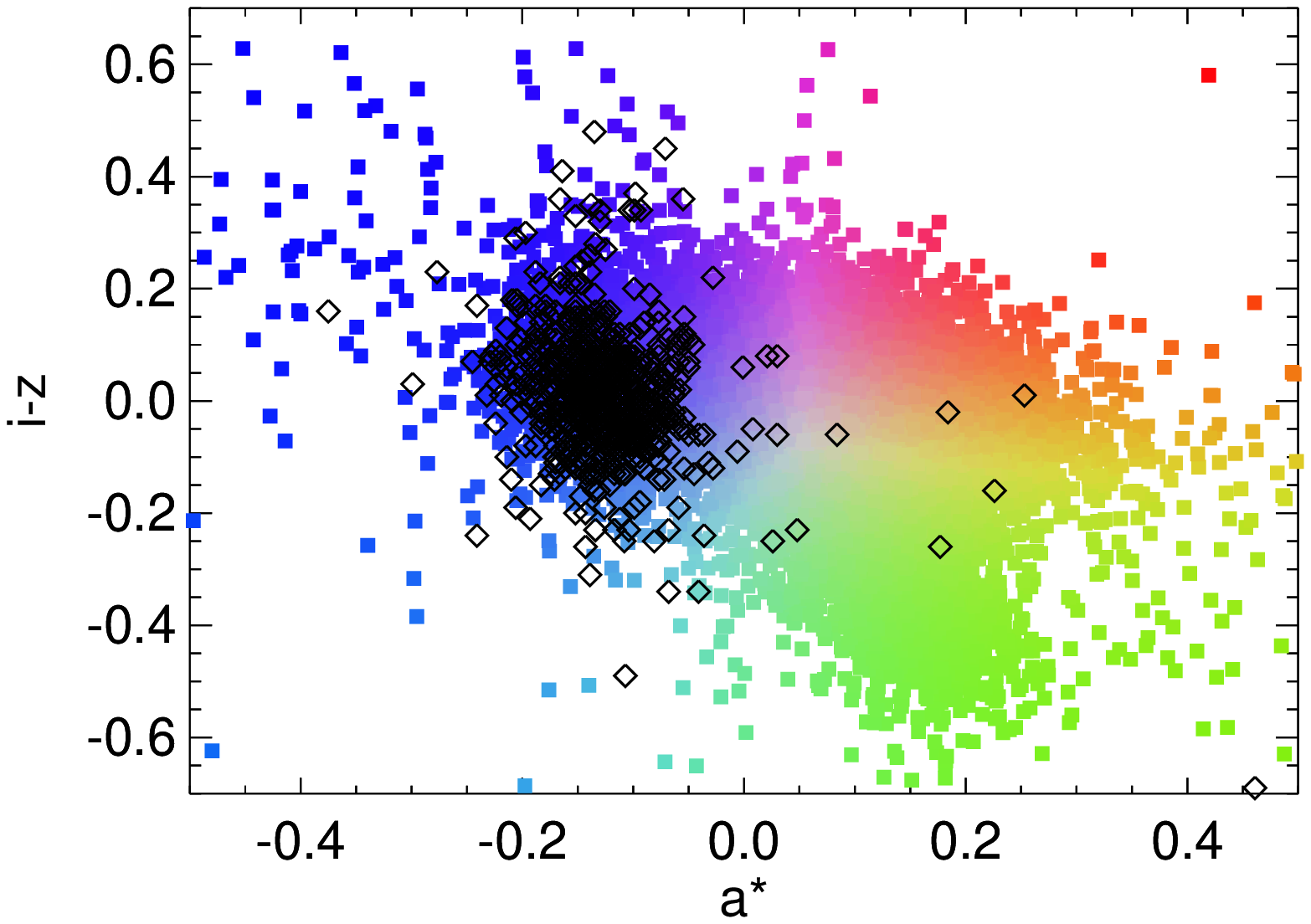}
\caption{The members of the Eulalia family (open diamonds;
  $C=10.5\times10^{-5}$~AU) and all the SDSS observed asteroids
  plotted as $i - z$ colors as a function of the slope $a^{\ast}$
  where the $i$ filter is centered at 769~nm, the $z$ at 925~nm and
  $a^\ast$ is the derived first principal component for the
  distribution of asteroid colors in the SDSS $r-i$ vs. $g-r$
  color-color diagram. The color scale is that used by Parker et
  al. (2008) that associates colors with locations on this color-color
  plot.
  \label{fig:SDSS}}
\end{figure}

Here we matched 668 asteroids from selected family bounded by the
$C=10.5\times10^{-5}$~AU that were
also measured by SDSS, and have plotted them on the same axes and
color scheme as Parker et al. (2008). As expected, the family is
tightly clustered in the region dominated by primitive asteroids
(Fig. \ref{fig:SDSS}). Without more significant quantitative analysis,
it is clear that cluster is actually much tighter than that found for
all primitive bodies. However, it is also clear that the cluster is
not tight enough to use SDSS colors as a tool to differentiate members
from different family associations.

\subsubsection{Known Spectra}

\begin{figure}[h!]
\includegraphics[angle=0,width=\linewidth]{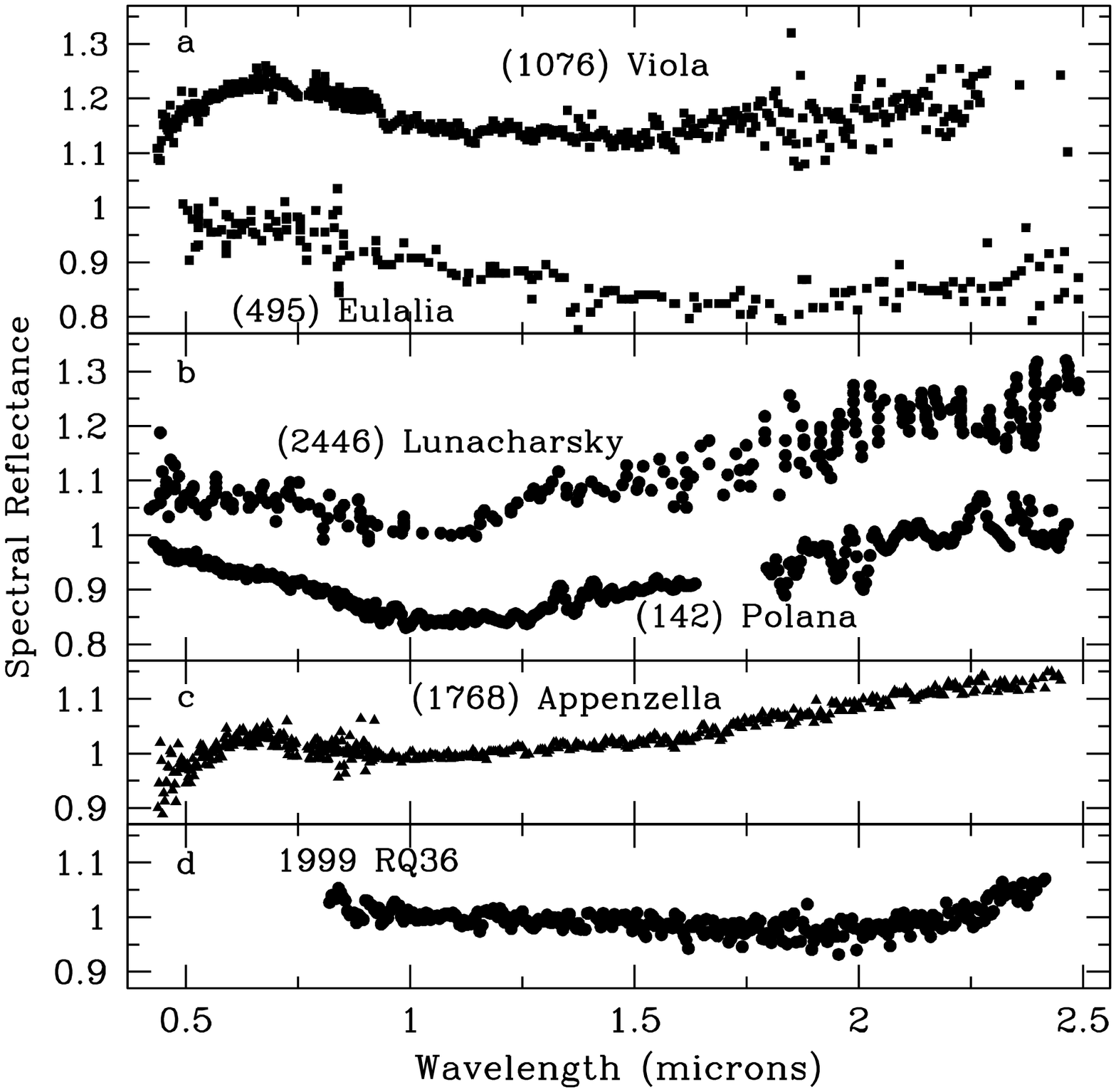}
\caption{Published visible and near-IR reflectance spectra of the two
  family parent bodies, (495) Eulalia and (142) Polana are plotted
  frames (a) and (b) respectively. The asteroid (1076) Viola is
  within the Yarkovsky lines for the Eulalia family (and outside the
  lines for the Polana family), and thus is plotted in frame (a) for
  comparison with (495) Eulalia. Asteroid (2446) Lunacharsky is within
  the Yarkovsky lines for the new Polana family and outside the lines
  for the Eulalia family, and is thus plotted in frame (b). Asteroid
  (1768) Appenzella is within the Yarkovsky lines for both families,
  and is plotted in frame (c). These spectra, due to the similarities
  within frames (a) and (b), hint that the family identification
  uncovered may indeed be two distinct families resolvable with
  visible and near-IR spectra. Finally, in frame (d) is a spectra of
  1999 RQ$_{36}$ for comparison (Clark et al. 2011).
\label{fig:SpecDiversity}}
\end{figure}

Asteroid (495) Eulalia was observed on 11~June~2007 at the NASA IRTF
with the SpeX instrument between 0.7--2.5 microns (Fieber-Beyer et
al. 2008; Fieber-Beyer 2011). There are no clear features for the
(495) Eulalia spectral reflectance, with only a minimal upturn in the
spectra beyond 2.0~microns (Fig. \ref{fig:SpecDiversity}). As the
parent of the family this spectra is potentially valuable in finding
differences between members of it and the neighboring
background. However, flat and featureless spectra are somewhat common
for low-albedo objects, which does not aid the determination of family
membership for an asteroid with unknown affiliation.

There exists numerous databases to search for spectra of other family
members, and two visible spectra (asteroids 1076 and 2509) were found among the
Small Main-belt Asteroid Spectroscopic Survey (SMASS; Xu et al. 1995,
1996) and two (asteroids 495 and 3999) in the second phase of that survey
(SMASSII; Bus and Binzel 2002). An additional two objects found in the
family using the $C=10.5\times10^{-5}$~AU boundaries (see
Fig. \ref{fig:TestAge10.5}) were observed in SMASS database (asteroids 1768 and
2809). No objects were found in the S3OS2 database (Lazzaro et
al. 2004) or in the PDS release of multiple IRTF surveys (Bus 2011).

The data in longer, near-IR, wavelengths is even more sparse. As
mentioned above (495) Eulalia, was observed with the IRTF SpeX
instrument as part of a campaign to study objects near the 3:1~MMR. As
part of the wider asteroid survey, the MIT-UH-IRTF Joint Campaign for
NEO Reconnaissance ({\tt http://smass.mit.edu/}), there are
near-infrared observations of asteroids (1076) Viola and (1768)
Appenzella. (1076) Viola, which belongs to the Eulalia family,
is plotted with (495) Eulalia (Fig. \ref{fig:SpecDiversity}) and shows similar morphology throughout, in particular they have a
similar slope beyond 2.0~microns. Asteroid (1768) Appenzella is within
the Yarkovsky boundaries for both families, and is thus plotted
alone. It shows an upturn starting around 1.0~micron, with a much
shallower slope than found for (142) Polana (discussed below).

\section{Discussion}\label{discussion}

An important aspect to the analysis of the Eulalia family is the
possibility of a larger, older family with similar orbits
and similar low albedo.  Here we investigate and characterize this proposed
family with some of the tools used above. This family is even more
extended and diffuse and much of the analysis is beyond the scope of
this current work.

\subsection{The ``new Polana'' family}
\label{background}

Beyond the clear correlation of the asteroids in the Eulalia family as
seen in Fig. \ref{fig:Vshape} and analyzed in the previous sections,
there appears to be a possible second family in the same dataset.  The
lack of asteroids with $a<$~2.3~AU and $H<$~14 is clear and defies
even the expected dropoff in the number of asteroids with decreasing
semimajor axis due to the inclination-dependent effects of the $\nu_6$
resonance and eccentricity limits of Mars crossing orbits. The high
density of objects with $15 < H < 17$ and $2.1$~AU $ < a < 2.3$~AU has
the characteristic ``ear'' of an asteroid family seen at higher
semimajor axis for the Eulalia family due to the Yarkovsky semimajor
axis drift.

This possible family suffers from the same, and even amplified,
difficulties found in analyzing the Eulalia family; a center near a
powerful resonance, overlapping with another family and possibly a
very old age. In many tests below we will remove the Eulalia family
from the data set (using previously defined boundaries) in an
attempt to measure only the possible background family.

\subsubsection{HCM of the background family}

Given the very diffuse nature of this possible family with an
unknown parent, we reverse the procedure used before, and begin
with an HCM test to make sure there is, in fact, an orbital
correlation among the high density region of asteroids. Given the very
low density of large objects, this HCM test was centered on the heart
of the visible ``ear'' of the family located at $2.2$~AU $ < a <
2.25$~AU and $15.5 < H < 16.5$. We simply selected a few objects in
this orbital range and ran the HCM at varying $V_c$ checking to see
whether any clump would acquire the characteristic shape in $H$
vs. $a$ space.

\begin{figure}[h!]
\includegraphics[width=\linewidth]{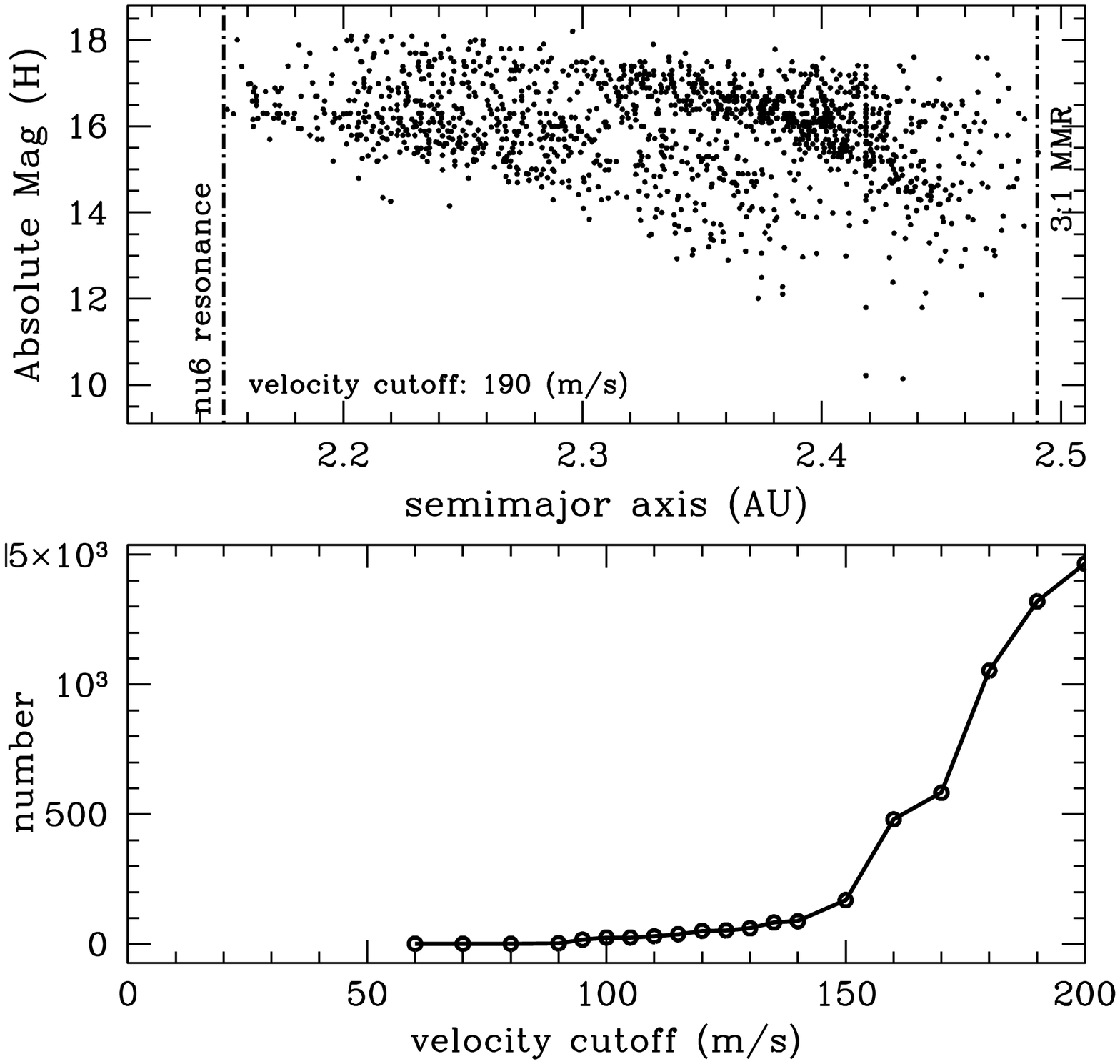}
\caption{The bottom panel shows the number of family members linked by
  the HCM routine to asteroid (2446) Lunacharsky identified as a
  likely member in the midst of the suspected new Polana
  family. The top panel shows the resulting HCM-linked family members
  for a $V_c = 190$~m/s, which shows the distinct sign of a Yarkovsky
  ``V-shape''. There is a clear over density of asteroids in the
  region of the Eulalia family, suggesting that our conservative
  attempt to remove the Eulalia family did not capture all family members.
  \label{fig:LunaHCM}}
\end{figure}

The first test was done with asteroid (2446) Lunacharsky. This
asteroid has proper orbital elements of $a=2.355$~AU, $e=0.154$,
$i=3.178^{\circ}$, and is also one of the few in this region with
published visible and near infrared spectra. The HCM results for this
body show that there is a highly diffuse and extended
association with few family members found below a velocity cutoff
of $V_c =$~150~m/s (Fig. \ref{fig:LunaHCM}, bottom). However, the association
linked by HCM at $V_c =$~180~m/s and larger shows the distinct signs in
the $H$~vs.~$a$ space of a correlated family and the family size
increases quickly.

\begin{figure}[h!]
\includegraphics[width=\linewidth]{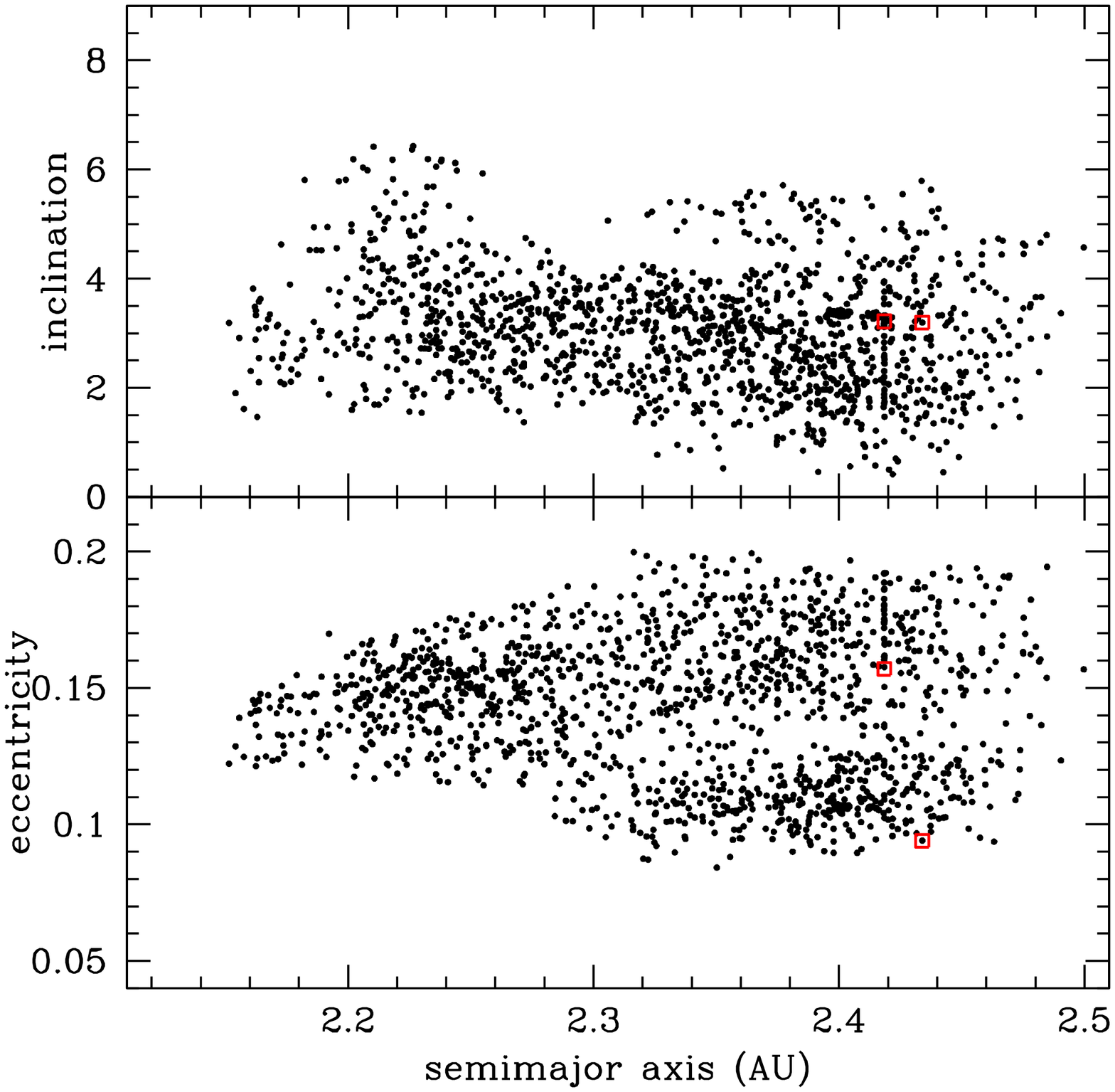}
\caption{The asteroids selected in the HCM search around (2446)
  Lunacharsky, which is identified as a likely member of the new
  Polana family. The asteroids are plotted as their inclination (top)
  and eccentricity (bottom) as a function of their semimajor
  axis. There are red boxes around asteroids (142) Polana and (112)
  Iphigenia, where the latter is at the lower eccentricity value.
  \label{fig:LunaAEI}}
\end{figure}

The largest bodies linked to this association are asteroids (142)
Polana and (112) Iphigenia with proper elements $a=2.418$~AU,
$e=0.157$, $i=3.215^{\circ}$ and $a=2.434$~AU, $e=0.094$,
$i=3.195^\circ$ respectively (Iphegenia is linked at $V_c =$~180~m/s
and Polana at $V_c =$~190~m/s). The orbital element distribution of
the selected asteroids is quite large in eccentricity, spanning $0.1 <
e < 0.2$, but more tightly clustered in inclination between
$1^\circ < i < 5^\circ$.  Each of these large asteroids are in
distinct clusters determined by their eccentricity, where (142) Polana
is in the midst of the high eccentricity grouping and (112) Iphigenia
with the smaller low eccentricity grouping (Fig. \ref{fig:LunaAEI}).

\subsubsection{The center and age of the new Polana family}

\begin{figure}[h!]
\includegraphics[width=\linewidth]{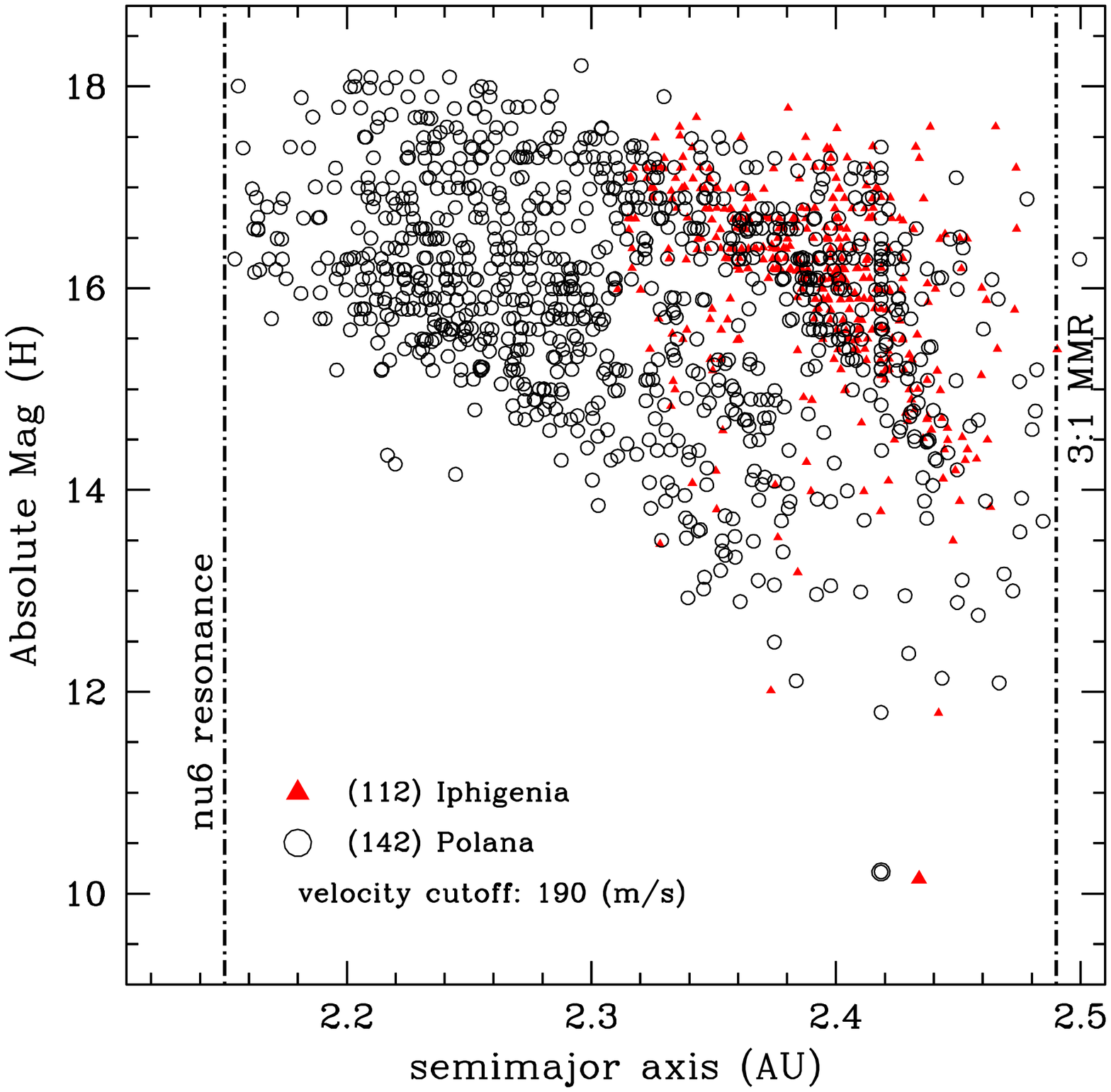}
\caption{The asteroids selected in the HCM search around (2446)
  Lunacharsky, which was identified as a likely member of the
  suspected new Polana family. The asteroids are plotted as
  their absolute magnitude as a function of semimajor axis, where the
  low-eccentricity cluster centered on (112) Iphigenia are plotted as
  red triangles, and those centered around (142) Polana are open black
  circles.
  \label{fig:PolSeparate}}
\end{figure}

We focus on (142) Polana and (112) Iphigenia due to their size and the
suggestions from the $H$ vs. $a$ plot that they are near the center of
the family. Differentiating them is difficult due to these
similarities, other than (142) Polana being embedded in a larger and
denser grouping in eccentricity. When these groups are separated and
plotted as $H$~vs.~$a$ it is clear that (142) Polana is associated
with the extended family structure
(Fig. \ref{fig:PolSeparate}). Meanwhile (112) Iphigenia is associated
with objects with larger $H$ and larger $a$, that are near the
highest-density region of the Eulalia family. It is possible that if
the Eulalia family was linked with a larger $V_c$ in the HCM routine
that many of these objects would be added to the Eulalia family.

\begin{figure}[h!]
\includegraphics[width=\linewidth]{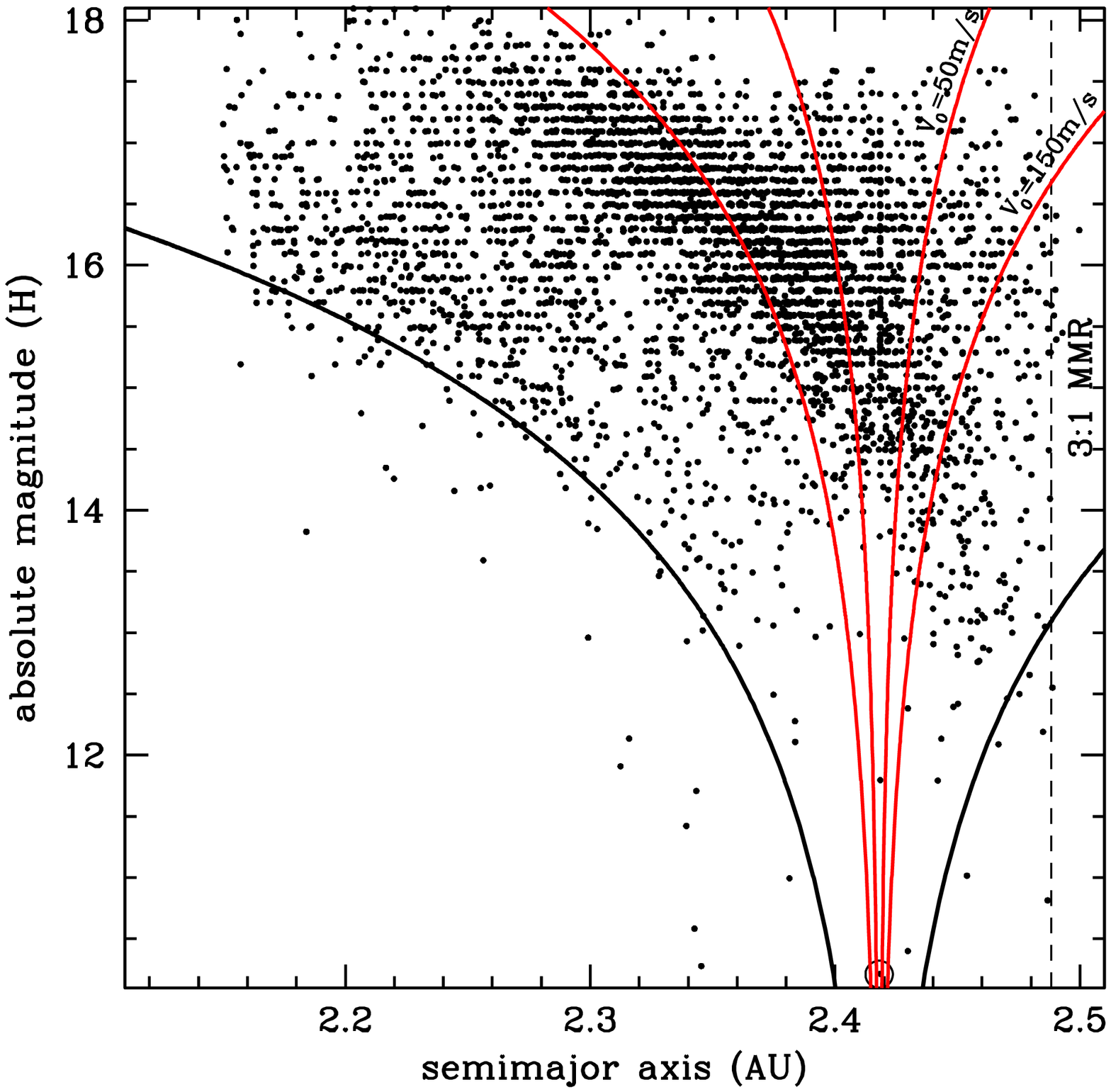}
\caption{Best fit boundary for the new Polana family. The Yarkovsky line
  for the family boundary at $C=16.9\times10^{-5}$~AU is the solid black
  line.  The red lines represents a simple initial dispersion for a
  velocity dispersion of fragments for $V_0=50$~m/s and $V_0=150$~m/s by
  $\Delta a=(2/n)V_0(5~\mathrm{km}/D)$ where $n$ is the mean
  motion. 
  \label{fig:Pol16.9}}
\end{figure}

Thus, the extended family here appears to be associated with asteroid
(142) Polana, and we use its proper semimajor axis $a=2.418$~AU as the
nominal center of the family. Using the same fitting routine as
before, we then measure a best-fit at $C=16.9\times10^{-5}$~AU
(Fig. \ref{fig:Pol16.9}). This fit was substantially easier due to the
dramatic density fall-off, and the measured best-fit $C$ value is
nearly twice as large as for the Eulalia family. This $C$ directly
correlates to $T=2845$~Myr, not accounting for the initial velocity
dispersion of the fragments following the collision. We can use the
same initial velocity dispersions of $V_0=50$~m/s and $V_0=150$~m/s,
despite not having a solid estimate of parent body size, 
which puts the age range of the Polana family at
$\sim$2000--2500~Myr. However, uncertainty
about YORP reorientation of spin axes adds age uncertainty which
becomes increasingly important at increasingly larger asteroid sizes for older
families. This adds substantial uncertainty to the upper boundary of
this age estimate, and the conservative estimate is simply an age
greater than 2000~Myr.

\subsubsection{Significance test of ``new Polana'' shape}

Given the low density of asteroids in the WISE dataset at low $a$ ($<
2.3$~AU) and $H$ ($< 13$), it is fair to question if the
characteristic shape of the new Polana family is not simply a
statistical fluke caused by decreasing asteroid density near the
$\nu_6$ secular resonance. Specificially, if the gaping hole in the
sample with $a < 2.3$~AU and $H < 13$ is due simply to statistical
flucuation, then the boundary of the new Polana Yarkovsky ``V-shape''
may simply be an artifact and not the edge of an asteroid family.

\begin{figure}[h!]
\includegraphics[width=\linewidth]{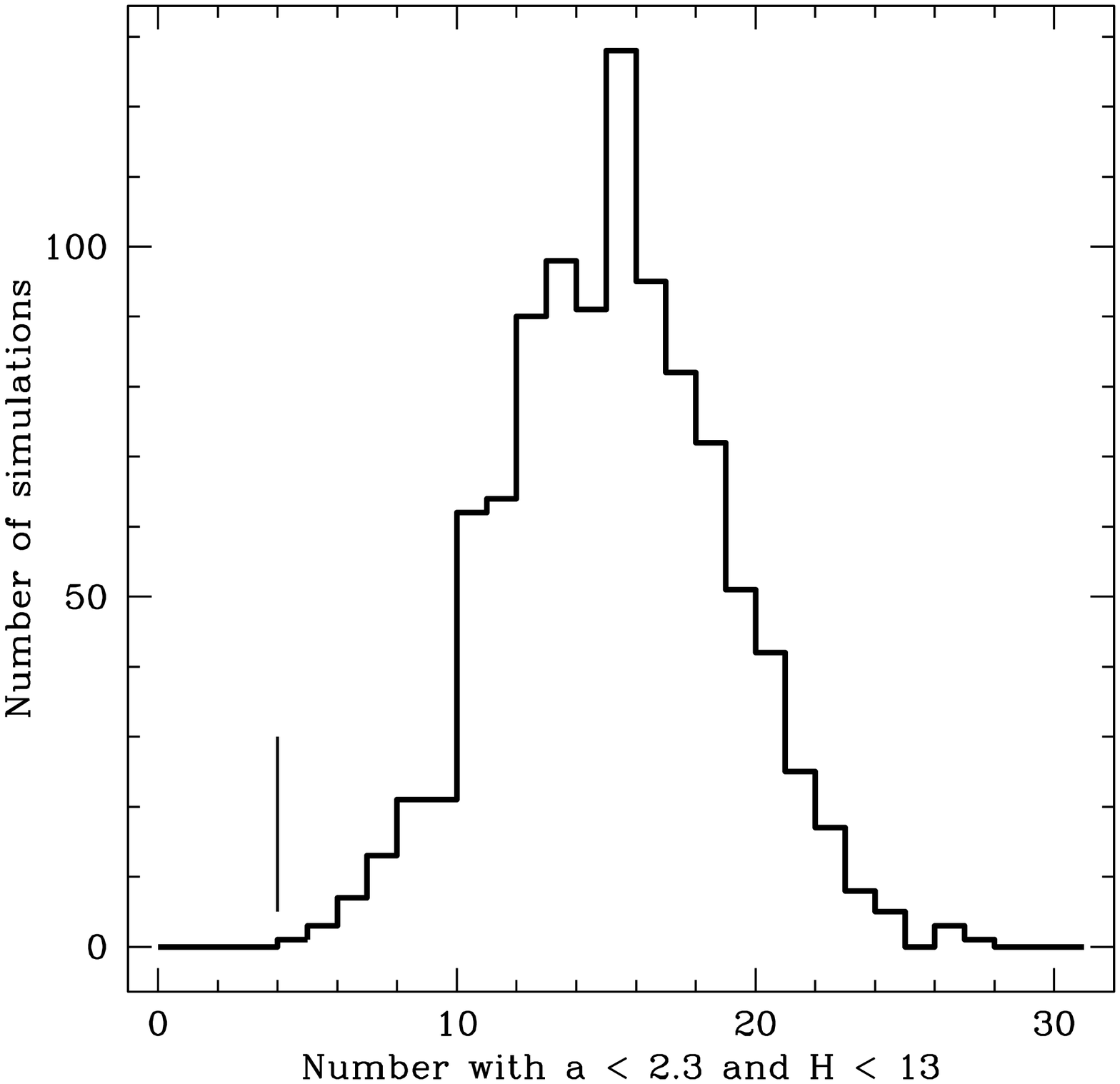}
\caption{The distribution of the number of asteroids with $a < 2.3$~AU
  and $H < 13$ for the 6702 asteroids (observed by WISE with \pv
    $< 0.1$) re-distributed with randomized $a$, $e$, $i$, and $H$
  distributions. The actual data had only 4 objects in this $a$
  vs. $H$ range (indicated by the vertical line) and is outside the bounds of the distribution found in
  the Monte Carlo test.
  \label{fig:fooled} }
\end{figure}

A simple test of the significance of this was to take the entire
distribution of asteroids in our dataset and randomly re-distribute them
while preserving the orbital and size distributions.  More simply, the
values of $a$, $e$, $i$ and $H$ were each separately randomized 1000
times.  A test of number of asteroids with $a < 2.3$~AU and $H < 13$
was performed for each. The WISE dataset has only 4 asteroids in this
region, while the average number found in this region was 15.7, and
the smallest number was 5 for all 1000 iterations (Fig. \ref{fig:fooled}).

This test did not specifically test if the V-shape was a fluke, but
rather whether the stark absence of asteroids at lower $H$ was
random. Finding that it is clearly significant suggests that the
density contrast seen is related to actual correlated family and
supports the existence of the new Polana family.

One alternative explanation for the lack of large low-albedo asteroids
inside of $\sim$~2.3~AU is simply that this is the innermost edge to
C-type asteroids in the asteroid belt. It is long known that primitive
bodies predominate in the outer asteroid belt (Gradie and Tedesco
1982), and some recent work has investigated various means to
``implant'' primitive bodies in the asteroid belt from more distant
regions of the Solar System (Levison et al. 2009; Walsh et
al. 2011). Levison et al. (2009) focus on capture of D- and P-type
asteroids during the Solar System instability associated with the Late
Heavy Bombardment, and they find a good match to the distribution of
these taxonomic classes, with the inner edge of both populations
around $\sim$~2.6~AU. Walsh et al. (2011) focused on the origin of the
C-class populations that are possibly scattered into the asteroid belt
much earlier in Solar System history. This work shows a steep decline
in primitive asteroid numbers at decreasing semimajor axis, and
theoretically could be responsible for the absence of large primitive
bodies inside of 2.3~AU.


\subsection{Spectral hints at two families}

The albedo distribution, SDSS colors and visible spectra of objects in
the family regions are very similar and not useful to distinguish
between families. Therefore any available spectra - especially in the
near-infrared - may be able to differentiate family membership.

There is publised near-IR spectra for one asteroid, (2446)
Lunacharsky (de Le\'{o}n et al. 2012), that could be a member of the
proposed Polana family, and it is plotted with (142) Polana, the
candidate parent (Fig. \ref{fig:SpecDiversity}, also see Fig. 4 in de
Le{\'o}n et al.  2012). Both objects show similar positive slopes
beyond 1.0--1.2~micron, similar negative slopes in the visible and
generally very similar morphologies.

Asteroid (495) Eulalia shows no clear feature at 1.0 or 1.2~micron in
contrast to (142) Polana.  There is also a clear slope
difference between the two family parents beyond 1.0~micron, which has
recently been studied as a diagnostic tool by de Le\'{o}n et
al. (2010) and Ziffer et al. (2011) to differentiate primitive
asteroids with otherwise very similar featureless visible spectra.
The hint of distinction between family parents, including additional
candidate family members, is tantalizing for the possibility that
near-infrared spectra could distinguish membership between the two
families for the numerous smaller bodies that fall within the
Yarkovsky boundaries of both. Similarly, this is enticing as a
possible tool to distinguish a possible origin for primitive NEOs that
have likely originated in this region of the asteroid belt, especially
future space-mission targets (for example, see panel {\bf d} of
  Figure \ref{fig:SpecDiversity} that displays the spectrum for NEO
  1999 RQ$_{36}$, the baseline target for NASA's OSIRIS-REx sample
  return mission).

\section{Conclusions and future work}

The structure and past evolution of the families in the inner Main
Belt are critically important in order to understand the history and
evolution of the asteroid belt as well as the delivery of asteroids to
NEO orbits. We have found that the primitive population in the inner
Main Belt is dominated by a few very old and large families. The
low-albedo component of the Nysa-Polana complex is an asteroid family
parented by asteroid (495) Eulalia. The family's age is between
900--1500~Myr, and the parent body had a size between 100--165~km.  We
have also found evidence for the existence of an older and larger
family centered on (142) Polana, dubbed here the ``new Polana''
family.  This family is parented by (142) Polana and formed more than
2000~Myr ago.

In light of this work, it appears that the major contributors of
primtive NEOs could be the Eulalia and new Polana family. Our
  dataset is limited to $H<18.8$, and therefore we cannot directly
  observe how the smallest members ($H>18.8$) of the family
  are behaving. We can, however, extrapolate based on the Yarkovsky
  curves, and estimate the $H$ value where the Yarkovsky curve crosses
  a major resonance. Asteroids with a larger $H$ can reach that
  resonance by simple Yarkovsky drift, while those with smaller $H$
  cannot. This method ignores complications due to YORP cycles, or
  other effects such as collisions, that may have a different
  size-dependence than the Yarkovsky effect.

In particular, we are interested in where the curves cross the $\nu_6$
resonance, as it is efficient at delivering NEOs.  The new Polana
family, being older, is delivering asteroids up to $H\sim 16$, while
the Eulalia family is only now contributing bodies of $H\sim
17.5$. Each family has also contributed a large fraction of their
total families into the 3:1~MMR. Most of these asteroids would have
been directly placed into the resonance immediately following the
family formation event, and therefore would have resulted in more of
an rapid influx, rather than the slow and measureable contribution via
the $\nu_6$ resonance.  Further dynamical models beyond the scope of this work
may also help to better discern the flux from each family to NEO
orbits over time via both the $\nu_6$ and 3:1~MMR.

One implication of the location of both families near the
  3:1~MMR, is that nearly all surviving family members have drifted
  inward from the family center over time. This inward drift requires
  a retrograde spin sense, which is a property that can be determined
  with lightcurve observations. It is possible to change the spin
  sense of an asteroid, by either collisions or YORP
  reorientation. Both families show an absence of small objects near
  the center of the family, suggesting that most have continued their
  inward drift without a substantial number suffering a spin-axis
  reorientation. This supports the implication that most members of
  both families should be retrograde rotators.

Each of the three proposed NEO space mission targets (1999 RQ$_{36}$,
1999 JU$_3$ and 1996 FG$_3$) are almost certainly ($>90$\%) delivered
from the inner region of the Main Asteroid belt, following the
well-studied dynamical pathway from the Main Belt to NEO-orbits
(Bottke et al. 2000,2002; Campins et al. 2010; Walsh et
al. 2012). More specifically, they appear to have come from the
$\nu_6$ resonance at 2.15~AU, which is the dominant supplier of NEOs
(Bottke et al. 2002), and their current low inclination is indicative
of similarly low inclination orbits in the Main Belt. Though entire
works are dedicated to understanding the origin of these specific
objects, the two families studied here are likely the dominant sources
of primitive NEOs. These conclusions have been reached previously in
works by Campins et al. (2010) finding the Nysa-Polana complex as
the likely source of 1999 RQ$_{36}$ (now probably from the Eulalia
family), or Gayon-Markt et al. (2012) finding that the background (now
largely belonging to the new Polana family) is an equal or greater
contributor of primitive NEOs.

A major part of this work was the description of relatively new
techniques to detect and analyse very diffuse and old asteroid
families. Despite significant previous work on these topics, the
discovery of both families was greatly aided by the use of the WISE
albedo data, which allowed specific analysis of low-albedo
objects. Beyond the advantage of focusing on asteroids with similar
physical properties, it also removed the ``noisy'' background of
high-albedo asteroids which otherwise would have convoluted the HCM
tests (resulting in the HCM linking the huge chunks of the asteroid belt). Thus
the slew of new data allows a pre-selected subset of the asteroid belt
from which to detect possible family correlations. 

The possible pitfalls of these technique are that the portion of the
asteroid belt selected could be biased or incomplete. Asteroids not
included in the subset of data, those not observed by WISE for
example, need to be tracked down in further iterations of the work to
build a full list of family members. Though this work attempted to
search for all possible parent members using numerous databases, (495)
Eulalia required special dynamical care to be recognised as the
parent, a task not realistic for a larger survey of very old
families.

Finally, there are potentially vast implications in assigning much of
what was considered to the be the background of the asteroid
  belt to an older family, the new Polana family. While this family
was particularly difficult to find, even while scouring a small subset
of the entire database of asteroids, it highlights what might still be
lurking in the data. How much of the background is actually part of
old and diffuse families?  If we continue with this exercise and
continue to discover old and diffuse families will there be any
background left, and what does this tell us about the origin of the
asteroid belt? Certainly, the answers to these questions are beyond
the scope of this project, but hopefully the flood of data from
current and future spacecraft (the ESA Gaia mission for example) and
larger ground-based surveys might help to answer these.

\section*{Acknowledgments}
KJW would like to thank NLSI for funding, and also the Gaia-GREAT
program of the European Science Foundation for funding a visit to
Nice. MD thanks the Centre National d'Etudes Spatiales (CNES) for
financial support.  DSL and WFB thank the NASA New Frontiers Program
for supporting this research (Contract NNM10AA11C). DV was supported
by the Grant Agency of the Czech Republic through grant
205/08/0064. We thank Alex Parker for the SDSS color palette.



\end{document}